\documentclass[aps,prx,floats,floatfix,twocolumn,fleqn,linenumbers]{revtex4-2}

\RequirePackage{lineno} 

\usepackage{graphicx}
\usepackage{epstopdf}
\usepackage{overpic}
\graphicspath{{./Figs/}{./}} 

\usepackage{latexsym}
\usepackage{amsmath}
\usepackage{amssymb}
\usepackage{bm}
\usepackage{wasysym}

\usepackage{ifthen}
\usepackage{color}
\usepackage[colorlinks,allcolors=blue,bookmarks=true]{hyperref}

\newcommand{\const}{\mbox{const}}

\newcommand{\mass}{\mathsf{m}}

\newcommand{\amatrix}[1]{\begin{matrix} #1 \end{matrix}}

\newcommand{\ket}[1]{\left| #1 \right\rangle}
\newcommand{\braket}[1]{\left\langle #1 \right\rangle }

\newcommand{\Braket}[2]{\left\langle #1 \middle| #2 \right\rangle}
\newcommand{\BraKet}[3]{\left\langle #1 \middle| #2 \middle| #3 \right\rangle}

\newcommand{\tr}{\mbox{\text{Tr}}}

\newcommand{\beq}{\begin{eqnarray}}
\newcommand{\eeq}{\end{eqnarray}}

\newcommand{\hide}[1]{}  
\newcommand{\rmrk}[1]{#1}

\newcommand{\Eq}[1]{{\textcolor{blue}{Eq.}}~\!\!(\ref{#1})} 

\newcommand{\App}[1]{\textcolor{blue}{{App}\!~(\ref{#1})}} 
\newcommand{\Fig}[1] {{\textcolor{blue}{Fig.}}~\!\!\ref{#1}}
\newcommand{\sect}[1]{{\bf #1.-- }}

\newcommand{\hrefl}[2]{\href{#2}{(#1)}}

\begin{document}


\title{Quantum irreversibility of quasistatic protocols \\ for finite-size quantized systems}

\author{Yehoshua Winsten, Doron Cohen (*)}

\affiliation{
\mbox{Department of Physics, Ben-Gurion University of the Negev, Beer-Sheva 84105, Israel} 
}

\begin{abstract}
Quantum mechanically, a driving process is expected to be reversible in the quasistatic limit, aka adiabatic theorem. This statement stands in opposition to classical mechanics, where \rmrk{mix of regular and chaotic dynamics} implies irreversibility. A paradigm for demonstrating the signatures of chaos in quantum irreversibility, is a sweep process whose objective is to transfer condensed bosons from a source orbital. \rmrk{We show that} such protocol is dominated by an interplay of adiabatic-shuttling and chaos-assisted depletion processes. The latter is implied by interaction-terms that spoil the Bogolyubov integrability of the Hamiltonian. As the sweep rate is lowered, a crossover to a regime that is dominated by quantum fluctuations is encountered, featuring a breakdown of quantum-to-classical correspondence. The major aspects of this picture are not captured by the common two-orbital approximation, which implies failure of the familiar manybody Landau-Zener paradigm.    
\end{abstract}

\maketitle

\section{Introduction} 

In Classical Mechanics, contrary to a prevailing misconception, the quasi-static limit is in general {\em not} adiabatic. This observation implies that protocols become irreversible, even if their control parameters are varied very very slowly. Adiabaticity and reversibility in the quasistatic limit are guaranteed only if the phase-space of the system does not undergo structural changes. Accordingly, one distinguishes between integrable-dynamics version of adiabaticity \cite{Landau} where action integrals serve as adiabatic invariants, and chaotic-dynamics version of adiabaticity \cite{Ott1,Ott2,Ott3,Wilkinson1,Wilkinson2,crs,frc} where the phase-space volume is the adiabatic invariant.  \rmrk{Generic systems feature {\em mixed phase-space} that contains both quasi-regular and chaotic dynamics.} Such systems do not obey the standard adiabatic theorems. The simplest demonstration for such irreversibility is the {\em seperatrix crossing} scenario that has been discussed extensively in the mathematical literature \cite{Kruskal,Neishtadt1,Timofeev,Henrard,Tennyson,Hannay,Cary,Neishtadt2,Elskens,Anglin,Neishtadt3}. But generic systems have more than a single degree-of-freedom, and therefore chaos becomes a central theme in the analysis \cite{Kedar1,Kedar2,apc,lbt,qtp}.

\begin{figure*}
\centering
\includegraphics[height=10cm]{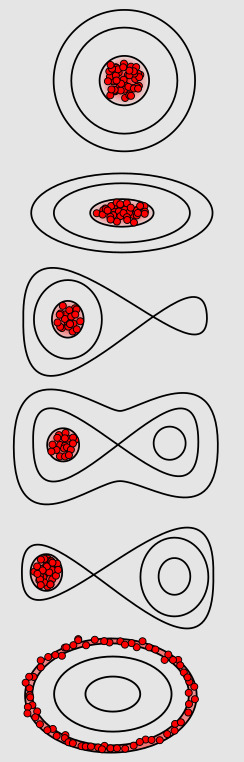}
\includegraphics[height=10cm]{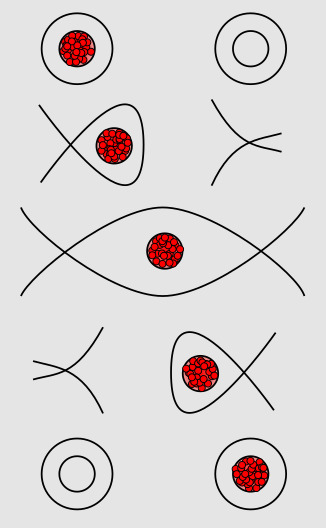}
\includegraphics[height=10cm]{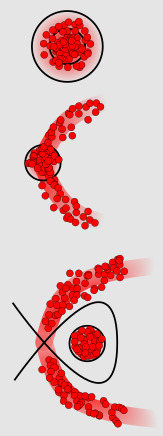}
\caption{ 
{\bf Schematics of phase-space evolution.} 
Each panel provides a sequence of phase-space snapshots.  
In the left and middle panels the curves are ${\mathcal{H}=\const}$ contours 
of a one-degree-freedom system. The evolving cloud is red. 
Initially the cloud is located in the minimum of the energy landscape.
The left panel displays an {\em adiabatic shuttling} process.  
As a control parameter is varied a second local minimum 
appears due to a saddle-node bifurcation (3rd snapshot), 
and the cloud becomes metastable (5th snapshot). 
In a quantum perspective the evolution is {\em diabatic}, 
meaning that quantum tunneling does not have the time to take place.
The process ends with {\em diabatic ejection} (last snapshot). 
If the sweep is reversed (not shown), 
the cloud can {\em split} into the two minima of the 5th snapshot
(assuming that both basins are expanding).          
The middle panel displays a {\em relay-shuttling} process. 
It consists of pitchfork bifurcation; 
swap of seperatrices;  
and inverse pitchfork bifurcation. 
The right panel displays the effect of 
a {\em chaos-assisted depletion} mechanism 
that competes with the pitchfork bifurcation 
of the relay-shuttling process.  
Strictly speaking we display in this panel 
a Poincare section of a two-degree-of-freedom system.
Due to spoiled integrability, 
there is a chaotic strip along which spreading is allowed.
The outer part of the cloud starts to spread away 
before the central fixed-point becomes unstable.  
}
\label{FigBifurcations}
\end{figure*}

In this paper we would like to explore how the above picture is reflected or modified {\em in the quantum framework}. The most suitable arena for such studies concern the dynamics of condensed bosons. In order to avoid an abstract discussion, let us consider a specific generic scenario. Let us assume that initially the bosons are condensed in a source orbital. A sweep protocol is designed to transfer them to a different orbital. {\em Naively}, one is inclined to speculate that this would be merely a many-body version of the Landau-Zener (LZ) adiabatic passage problem. The classical limit, aka nonlinear LZ problem, has been studied extensively \cite{LZ-Niu-PRL,LZ-Niu-PRA}. It features a {\em diabatic ejection} stage (\Fig{FigBifurcations}, left panel) that is related to a swallow-tail structure in its bifurcation diagram. The full quantum version has been addressed as well \cite{cst}. Irreversibility has not been discussed  there, but it is expected due to the seperatrix crossing, per the conditions of the Kruskal-Neishtadt-Henrard theorem \cite{Kruskal,Neishtadt1,Timofeev,Henrard,Tennyson,Hannay,Cary,Neishtadt2,Elskens,Anglin,Neishtadt3}.

We claim that in general the manybody LZ problem cannot serve as a paradigm for depletion. Typically the dynamics involves more than two orbitals, meaning that we are dealing with more than one degree of freedom. Consequently the role of {\em chaos} cannot be ignored \cite{apc,lbt,qtp}. Using different phrasing, we say that the inapplicability of the LZ paradigm is related to the failure of the two-orbital approximation (TOA). Once additional orbitals are taken into account, the integrability of the Hamiltonian is spoiled. Consequently, the depletion stage involves competing mechanisms which we call {\em adiabatic shuttling} and {\em chaos-assisted depletion} (\Fig{FigBifurcations}, middle and right panles).

Our interest is to address the {\em irreversibility} theme, and to contrast {\em quantum} against {\em semiclassical} dynamics. In our semantics the term `semiclassical' replaces the term `classical' whenever the quantum state is represented in phase-space by a cloud of points, that are propagated using classical equations of motion. This is also known as the `truncated-Wigner-approximation', and goes much beyond the single-trajectory dynamics of Mean Field theory. Nevertheless, semiclassical approximation, in this restricted sense, is not capable of taking into account neither tunneling \cite{dimerSplit,tunnleKAM,tunnleHeller,apq} nor interference of separated trajectories.

In quantum mechanics, contrary to the semiclassical picture, the quasi-static limit of a closed finite system is always adiabatic, and therefore reversible. This is because the energies are quantized, and therefore the system follows the (gaped) ground state for slow enough driving. However, this quantum adiabaticity has no experimental significance once we deal with a mesoscopic system. 
In the example that we discuss in this work, the condensate is a flow-state of a superfluid ring. As the control parameter is varied, the flow-state becomes {\em metastable}. But the tunnel coupling to the new ground state is exponentially small in the number of particles \cite{dimerSplit}, and therefore can be ignored. Hence the system fails to follow the ground state. This is in fact the essence of superfluidity. The question remains, what is the fate of the flow-state as the control parameter is further varied. What is the mechanism of depletion? Do we have the same irreversibility as in the semiclassical analysis?

The question that we pose is not merely related to the foundations of physics (irreversibility, quantum vs classical). It is also of practical importance for the design of protocols whose objective is to manipulate manybody states of cold atoms, aka atomtronics \cite{atomtronics}.
Specifically, we consider bosons that are described by the Bose-Hubbard Hamiltonian (BHH). This model is of major interest both theoretically and experimentally \cite{Oberthaler,Steinhauer,exprBHH1,exprBHH2}. There is a particular interest in lattice-ring circuits that can serve as a SQUID or as a useful Qubit device \cite{Amico,Paraoanu,Hallwood,sfr}. The hope is to achieve coherent operation for BHH configurations that involve a few orbitals. 
This is the natural extension of studies that concern two orbitals, aka Bosonic Josephson Junction. The most promising configuration is naturally the 3-site trimer  
\cite{ref12,trimer2,trimer3,trimer4,trimer6,trimer15,trimer7,trimer19,trimer8,trimer20,trimer18,trimer12,trimer13,trimerSREP2,gallemi,sfs,sfc,sfa,bhm}. 
For the analysis of such circuits one has to confront the handling of an underlying mixed phase space \cite{KolovskyReview,sfc,sfa}.

We are inspired by hysteresis experiments, as done for double well geometry \cite{exprDimerHys}, and by protocols that have been realized experimentally for bosons in a ring (or SQUID) geometry \cite{exprRingRev,exprRingNIST,exprRingLANL1,exprRingLANL2,atomtronics}. 
The related theoretical studies adopt the TOA, and highlight 
the appearance of swallow-tail bifurcations \cite{Swallow1,Swallow2,Swallow3,SwallowBrand,Swallow5}. 
\rmrk{But the failure of the TOA is anticipated} by observing that the Bogolyubov pairing interaction requires 3 orbitals, and by the further observation that there are additional terms in the Hamiltonian that spoils the integrability of the Bogolyubov approximation.  
\rmrk{Consequently,} our interest below is to push the discussion of irreversibility into the realm of high-dimensional dynamics, addressing the fingerprints of chaos and mixed phase-space in the quantum-mechanical reality.

\rmrk{The classical analysis of the forward sweep process follows our previous publication \cite{qtp}.}
In the present paper we further illuminate that the integrable mechanism that is implied by the Bogolyubov approximation is a variant of adiabatic shuttling that we call {\em relay shuttling} (\Fig{FigBifurcations}). In the quasistatic limit this mechanism is overwhelmed by chaos-assisted depletion. We explore the {\em quantum} scenario, and append an {\em inverse-sweep} of the control parameter, in order to study the {\em irreversibility} due to the interplay of the various mechanisms involved.
\rmrk{Our major observation is the discovery of a novel regime of {\em quantum irreversibility}, that has not been anticipated by the semiclassical analysis of \cite{qtp}.  This new regime features {\em universal quantum fluctuations}~(UQF), and an unexpected breakdown of {\em quantum-to-classical correspondence}~(QCC).} 
\\

\sect{Outline}
We present the Bose-Hubbard Hamiltonian that describes a superfluid ring, 
and display some results of simulations that probe irreversibility. 
\rmrk{The protocol for a proposed experiment with atomtronic circuit is highlighted: 
a superfluid ring whose rotation velocity is gradually increased and then decreased back to zero.}
We illuminate our findings by performing step-by step analysis: 
We clarify the failure of the TOA; 
we provide predictions that are based on the Bogolyubov approximation; 
and then, going beyond that, we discuss the implications of chaos. 
This is followed by a discussion, where we highlight the manifestation 
of UQF and the breakdown of QCC.      
\\ \ \\

\section{The model}

Consider $N$ bosons in an $L$-site ring, 
described by the Bose-Hubbard Hamiltonian (BHH)
with hopping frequency~$K$ and on-site interaction~$U$.
The sweep control-parameter is the Sagnac phase~$\Phi$, 
which is proportional to the rotation velocity $\Omega$ of the device.  
This phase can be regarded as the Aharonov-Bohm flux that is associated
with a Coriolis field in the rotating frame.
The Hamiltonian is 
\beq \label{eHa}
\mathcal{H} \ &=& \ \sum_{j=0}^{L-1} 
\Big[ 
\epsilon_j a^{\dag}_j a_j 
+ \frac{U}{2} \left( a^{\dag}_j a^{\dag}_j a_j a_j \right)   
\nonumber \\ &&
- \frac{K}{2} \left( e^{i\frac{\Phi}{L}} a^{\dag}_{j+1} a_j + e^{-i\frac{\Phi}{L}} a^{\dag}_j a_{j+1} \right)
\Big]
\eeq
where ${ \epsilon_j = -\epsilon\cos(2\pi j/L) }$ is included, as in \cite{SwallowBrand}. 
It signifies an external gravitation potential that may arise due to an optional \rmrk{{\em tilt}} of the ring. Some optional representations of the Hamiltonian are presented in \App{ApB} and \App{ApC}. 
Unless stated otherwise we assume $\epsilon{=}0$.
The notation ${u=NU/K}$ stands for the dimensionless interaction strength, 
and in the numerical simulations we use units of time such that $K{=}1$.

The momentum orbitals are labeled by the wavenumber ${k=(2\pi/L) \times \text{integer}}$, 
where the integer is the winding number. In this basis the Hamiltonian takes the form
\beq \label{eHb}
\mathcal{H} \ &=& \ 
\sum_{k=0}^{L-1}  \mathcal{E}_k b^{\dag}_kb_k  
\ - \ \frac{\epsilon}{2} \sum_{k,\pm}  b^{\dag}_{k \pm 1} b_{k} 
\nonumber \\
&&\ + \ \frac{U}{2L} \sum_{k_1,k_2,k_3,k_4}^{'} b^{\dag}_{k_1}b^{\dag}_{k_2}b_{k_3}b_{k_4} 
\eeq
where the prime in the $k$ summation implies that conservation of total momentum is required.
The presence of the control parameter $\Phi$ is implicit via  
\beq 
\mathcal{E}_k \ = \ -K \cos\left(k- \frac{\Phi}{L} \right)
\eeq

\sect{Preparation}
We start with a non rotating ring ($\Phi{=}0$). 
Initially the bosons are condensed in the 
zero momentum orbital ($k_0{=}0$).     
Keeping only the 3 lowest orbitals, 
labeled as  ${ (k_0,k_{+},k_{-}) }$,  
it is convenient to describe their subsequent occupation 
using the depletion coordinate~$n$, 
and the imbalance coordinate~$M$, that are defined as follows:
\beq
n \ \ &=& \ \ \sum_{k\ne0} n_k \ \ = \ \ n_{+}+n_{-} \\
M \ \ &=& \ \ n_{+} - n_{-}  
\eeq

\sect{Quantum chaos}
\rmrk{One can regard the BHH as the Hamiltonian of coupled non-linear oscillators. Standard analysis reveals that the underlying classical phase space is a mix of chaotic and quasi-regular regions. This may have signatures both in the many-body eigenstates that are labelled using a running index $\nu$, and in the statistics of the associated eigenenergies $E_{\nu}$. Respectively, one can characterize the spectrum using ``quantum chaos" measures $s$ and $r$, see \App{AppSgntr}. Such type of characterization has been illustrated e.g. in Fig.1 of \cite{bhe} for a trimer chain. A more refined version of~$s$, and discussion of its $L$~dependence has been provided in \cite{sgntr}.} 

\begin{figure}
\centering
\includegraphics[width=7cm]{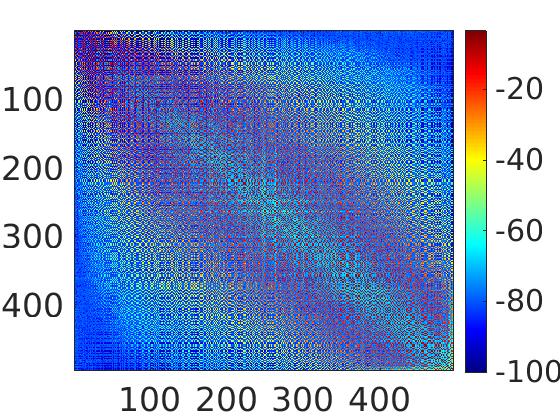} \\
\includegraphics[width=8cm]{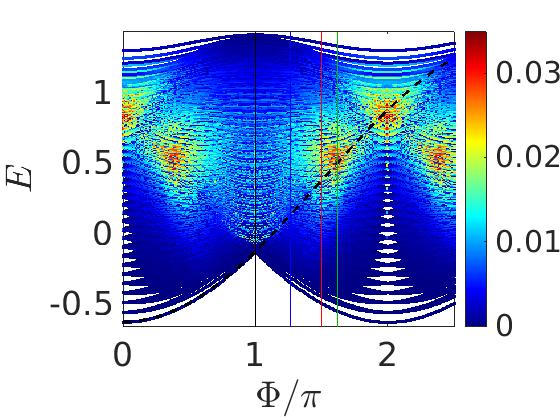}
\caption{ 
{\bf \rmrk{Signatures of quantum chaos.}} 
We consider  a trimer with $N{=}30$ particles. 
The upper panel is an image of the matrix $|I_{\nu,\mu}|^2$, 
color-coded in log scale.  
The model parameters are $K{=}1$ and $u{=}2.3$ and $\Phi{=}1.6\pi$.  
From this matrix we extract the chaoticity measure $s_{\nu}$ for each energy level. 
The lower panel shows the energy levels $E_{\nu}(\Phi)$ as a function of $\Phi \in [0,2.5\pi]$. 
The levels are color-coded by~$s$.   
The black line indicates the energy of the $k{=}0$ condensate, if it is not depleted.  
The vertical lines are the thresholds 
$\Phi_{\text{mts}}{=}1\pi$ (black), 
$\Phi_{\text{stb}}{=}1.26\pi$ (blue), 
$\Phi_{\text{dyn}}{=}1.5\pi$ (red), 
$\Phi_{\text{swp}}{=}1.62\pi$ (green). 
Later we we shall see that in a sweep process a depletion process takes place
during the time that $\Phi$ crosses these thresholds.      
}
\label{FigSgntr}
\end{figure}

\rmrk{In the present context the $r$ indicator is not useful, because we have mixed phase-space, and the chaotic region of interest is rather small. In contrast, the $s$ indicator is informative. 
\Fig{FigSgntr} provides an illustration of the matrix 
${ I_{\nu,\mu} = \BraKet{\nu}{(-\partial \mathcal{H}/ \partial \Phi)}{\mu} }$.
The band-profile of this matrix is related to the correlator of the current operator~$I(t)$. 
The quantum chaos indicator $s_{\nu}$ is extracted from this matrix for each energy level.  
A second panel displays the variation of the energy levels versus the control parameter $\Phi$. 
The levels are color-coded by $s$. Vertical lines indicate the thresholds $\Phi_{\text{mts}}$ (black), $\Phi_{\text{stb}}$ (blue), $\Phi_{\text{dyn}}$ (red), and $\Phi_{\text{swp}}$ (green). The first threshold ${\Phi_{\text{mts}}}$ is positioned where the $k_{+}$ orbital crosses the $k_0$ orbital and becomes the lowest in energy. The other thresholds will be defined in later sections, namely,  at $\Phi_{\text{stb}}$ the Landau stability is lost; at $\Phi_{\text{dyn}}$ dynamical stability is lost; and at $\Phi_{\text{swp}}$ we have a swap of seperatrices that is related to the relay shuttling mechanism.}

\begin{figure*}
\centering
\includegraphics[width=8cm]{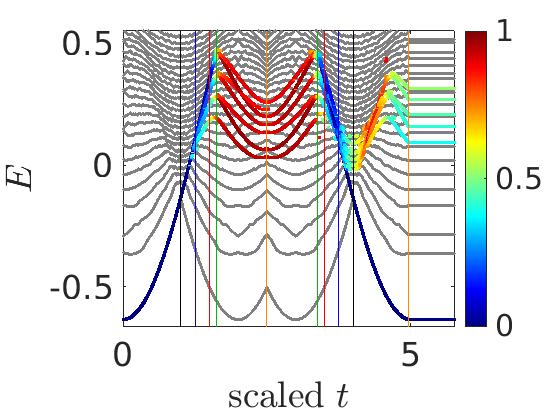}
\includegraphics[width=8cm]{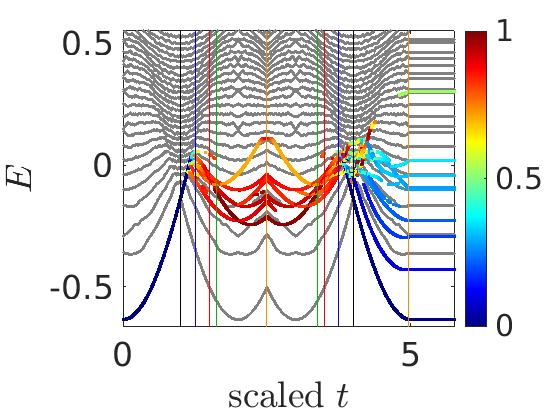} \\
\includegraphics[width=8cm]{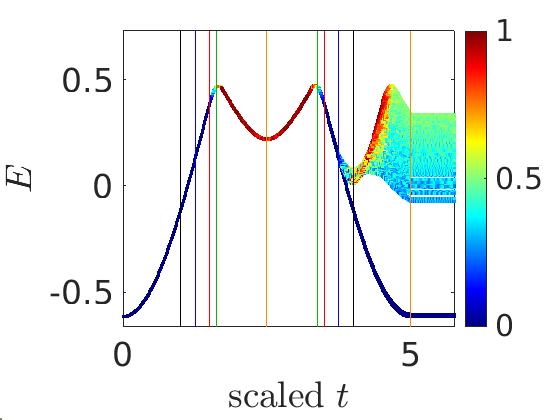}
\includegraphics[width=8cm]{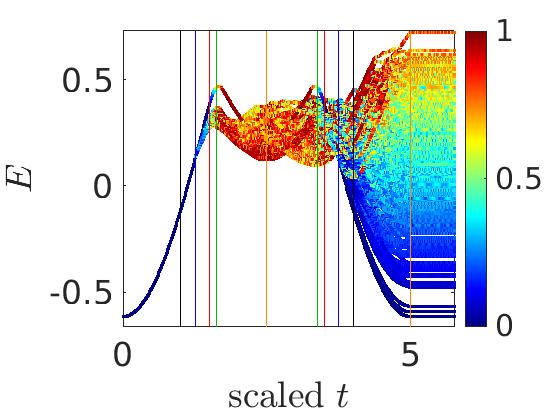} \\ 
\includegraphics[width=8cm]{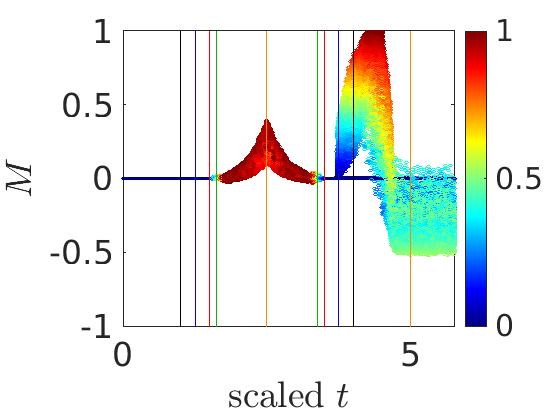}
\includegraphics[width=8cm]{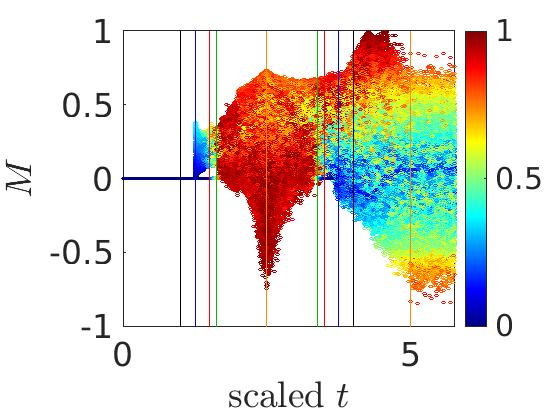}

\caption{ 
{\bf Simulations that test irreversibility.} 
The control parameter is swept from $\Phi{=}0$ to $\Phi{=}2.5\pi$ and back to $\Phi{=}0$.    
The horizontal axis is the scaled time ${ (1/\pi)|\dot{\Phi}| t }$. 
The vertical lines are as in \Fig{FigSgntr}.
The two additional orange lines indicate where the sweep is reversed ($\Phi{=}2.5\pi$) and stopped (back at $\Phi{=}0$). 
Note that the quantum simulations may include an additional waiting period at $\Phi{=}2.5\pi$. 
The initial state is the ground-state condensate.  
The upper panels are quantum simulation for $N{=}30$ particles, with $K{=}1$ and $u{=}2.3$.  
The energy levels $E_{\nu}(\Phi(t))$ are plotted.
Gray color indicates levels whose weight $p_{\nu}$ is vanishingly small  
(for presentation purpose this set is diluted by factor~10).      
The participation levels whose $p_{\nu}$  is non-negligible are color-coded by ${ \braket{n}_{\nu} }$.     
The left panel is for $\dot{\Phi}{=}5\pi \cdot 10^{-4}$, 
and the right panel is for $\dot{\Phi}{=}3.33\pi \cdot 10^{-7}$. 
The 2nd and 3r rows display semicalssical simulations for the same system. 
We plot $E$ and $M$ for an ensemble of trajectories,  
starting with a cloud that mimics the initial condensate. 
The value of the $n$-coordinate is color-coded. 
The left panel is for a slow sweep $\dot{\Phi}{=}5\pi \cdot 10^{-4}$, 
while the right is for a {\em very} slow sweep $\dot{\Phi}{=}5\pi \cdot 10^{-5}$. 
In this figure, and in all subsequent figures, 
the units are normalized (${n:=n/N, M:=M/N, E:=E/N}$).  
}
\label{FigTrimerEnergyVsTimeQM}
\label{FigTrimerEnergyVsTimeSC}
\end{figure*}

\rmrk{\section{Probing irreversibility using an atomtronic circuit}}

\subsection{The proposed experimental setup}

Consider a ring with condensed bosons. The optical potential that holds the bosons is possibly painted as in \cite{exprRingLANL1}. The ring has several weak links (as in SQUID geometry), or it can be an $L$-site lattice ring (as assumed below). Initially the ring is at rest, and the condensed bosons have zero momentum. In a {\em quench} protocol the ring starts abruptly to rotate. Superfluidity means that the rotation velocity $\Omega$ should be larger than a critical value $\Omega_c$ in order to induce current. The appearance of a non-zero current (depletion of the zero momentum orbital) can be verified using a standard time-of-flight measurement procedure. We would like to consider a {\em sweep} protocol, such that the rotation velocity is increased gradually (quasi-statically) from zero to a finite value that is larger than $\Omega_c$. Then we ask whether this sweep process is reversible. Accordingly, we decrease gradually $\Omega$ back to zero. Our main message, from the perspective of an experiment, is that the quasi-static protocol features novel {\em quantum irreversibility}. A secondary message is that the value of $\Omega_c$ is affected by the sweep rate, and provides an indication for the underlying depletion mechanism.


\subsection{Results of numerical simulations}

We present some results of numerical simulation for an $L{=}3$ ring, aka trimer. 
This will motivate the analysis in the subsequent sections.   
Initially all the particles are condensed in $k{=}0$, 
meaning that the initial value of the depletion coordinate is $n{=}0$.
The protocol consist of 3 stages: 
a forward sweep of $\Phi$ from $\Phi{=}0$ up to $\Phi{=}2.5\pi$, 
an optional waiting period, and a backward sweep to $\Phi{=}0$. 
Note that once $\Phi$ exceeds ${\Phi_{\text{mts}}=\pi}$ 
(to be indicated by black vertical line in the time axis of our figures)     
the condensate becomes metastable. But its depletion happens only 
in a later stage, as discussed below.

We display in \Fig{FigTrimerEnergyVsTimeSC} the variation of $(E,n)$ as a function of time 
using both quantum and semiclassical simulations. The variation of $n$ is color coded. 
In the semiclassical simulations we propagate an ensemble of trajectories, 
starting with a cloud that mimics the initial condensate. 
In the quantum simulations we propagate the evolving manybody state $\Psi(t)$, 
and calculate the probabilities 
\beq
p_{\nu}(t) \ \ = \ \ \left| \Braket{E_{\nu}}{\Psi(t)} \right|^2 
\eeq 
The energy levels $E_{\nu}(\Phi(t))$ are plotted as a function of time: 
gray color indicates levels whose weight is vanishingly small (less than 3.5\%); 
and the other levels whose $p_{\nu}$ is non-negligible are color-coded 
by ${ \braket{n}_{\nu}=\BraKet{E_{\nu}}{n}{E_{\nu}} }$.

One observes that for "slow" sweep the spreading in~$E$ is worse, 
indicating that irreversibility is enhanced. 
For the semiclassical simulation we show in \Fig{FigTrimerEnergyVsTimeSC} (3rd row) 
how this spreading is expressed in~$M$. 
The optional \Fig{FigTrimerDynamicsSplitting} of \App{ApBranching} 
shows how the spreading looks like in occupation space, 
using ${(n,M)}$ coordinates.

\begin{figure}
\centering
\includegraphics[width=8cm]{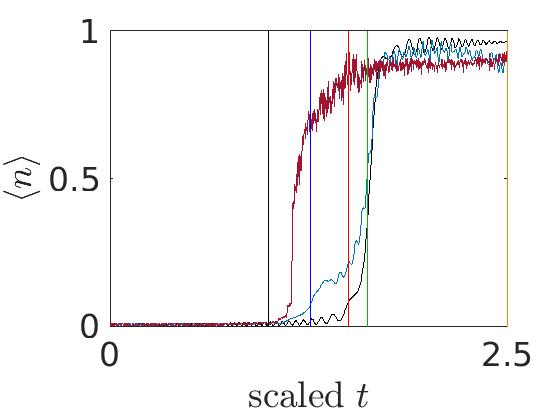} \\
\includegraphics[width=8cm]{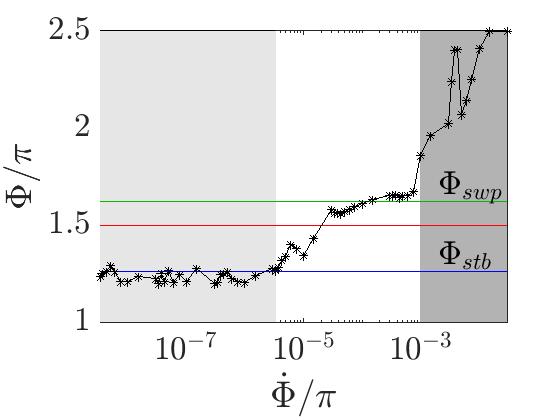}
\caption{ 
{\bf Depletion vs sweep-rate.} 
The upper panel displays the depletion $\braket{n}$ versus time for a quantum simulation 
with sweep rate $\dot{\Phi}{=}5\pi \cdot 10^{-4}$ (blue), 
and with very slow rate $\dot{\Phi}{=}3.33\pi \cdot 10^{-7}$ (red).
The former is compared with simulation (black) 
that is generated by the Bogolyubov-approximated Hamiltonian.  
The vertical lines and the parameters are as in \Fig{FigTrimerEnergyVsTimeQM}. 
From such plots we determine the time $t_d$ at which the depletion happens. 
The dependence of $\Phi(t_d)$  on the sweep rate $\dot{\Phi}$ 
is displayed in the lower panel. The dark gray background indicates non-quasistatic regime 
where the depletion time lags and becomes numerically ill-defined.  
In the quasistatic regime the observed dependence on $\dot{\Phi}$ indicates  
the crossover from chaos-assisted depletion (light gray background) to adiabatic shuttling.
Namely, the depletion shifts from $\Phi_{\text{stb}}$ to $\Phi_{\text{swp}}$.   
}
\label{FigTrimerDepletionTimeVsRate}
\end{figure}

\begin{figure}
\centering
\includegraphics[width=8cm]{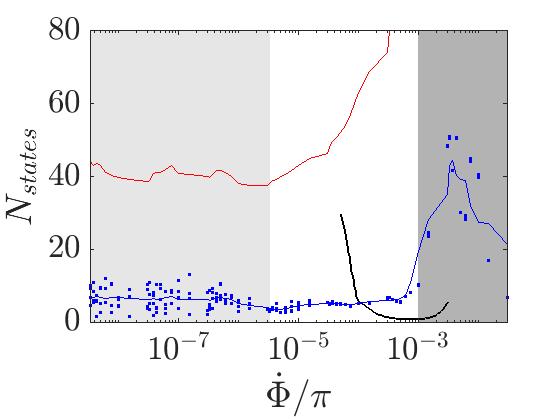} \\
\includegraphics[width=8cm]{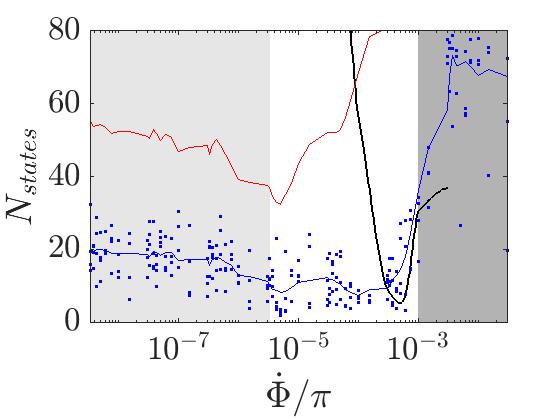}

\caption{ 
{\bf Irreversibility vs sweep-rate.}
Irreversibility is indicated by the growth of the the number $N_{\text{states}}$ 
of energy levels that participate in the evolution.     
We show the dependence of  $N_{\text{states}}$  (blue dots) 
on the sweep rate $\dot{\Phi}$ before the reversed sweep (upper panel) 
and at the end of the reversed sweep (lower panel),
for misc values of the waiting time.  
The erratic dependence on the waiting is illustrated in \Fig{FigTrimerSpreadVsWaiting-App}  of \App{ApN}.
The blue lines provides the average value of $N_{\text{states}}$, 
and the red lines provided the average value ${ \sum \nu p_{\nu} }$. 
%
The black lines are based on the semiclassical simulations. 
The gray background is the same as in \Fig{FigTrimerDepletionTimeVsRate}.
}
\label{FigTrimerSpreadVsRate}
\end{figure}

In \Fig{FigTrimerDepletionTimeVsRate} we plot the depletion $\braket{n}$ versus time.
In the quasistatic regime the time of the depletion $t_d$ is determined 
by inspection of the sharp rise in $\braket{n}$. 
We indicate by dark gray background color 
the range of $\dot{\Phi}$ where $t_d$ becomes ill-defined, 
reflecting a lag with respect to the parametric variation of~$\Phi$.       
In the quasistatic regime we observe that $\Phi(t_d)$ is shifted 
as $\dot{\Phi}$ is increased. Later we interpret this shift 
as an indication for a crossover from chaos-assisted depletion to adiabatic shuttling.

In order to quantify the {\em adiabaticity} in the quantum simulations,  
we characterize the spreading in energy by estimating the number 
of participating energy levels  
\beq
N_{\text{states}}(t) \ = \ \left[ \sum_{\nu}  \left| p_{\nu}(t) \right|^2 \right]^{-1}
\eeq
An optional measure is $N_{\text{orbitals}}(t)$ of \App{ApA}.
Illustrations for the temporal variation of both measures are provided in \App{ApN}.
It should be noted that $N_{\text{states}}(t)$ is expected 
to be monotonic increasing only for a strictly quasistatic process, 
which is not the case here (because we have mixed phase space and bifurcations along the way). 
Nevertheless, the final spreading can be used as a measure for the {\em irreversibility} of the sweep protocol.  
Its dependence on the rate $\dot{\Phi}$ is displayed in \Fig{FigTrimerSpreadVsRate}.

\hide{In plots that describe evolution versus time, 
we indicate by sequence of 4~vertical lines (black, blue, red, green) the range of $\Phi$ 
where non-adiabatic transitions are expected.  
A detailed analysis will be presented in subsequent paragraphs.
But for now we would like to make a few observations.}
%
We see that in the quasistatic regime slowness is bad for adiabaticity. 
This is very pronounced in the semiclassical simulation,
and has modest reflection in the quantum evolution. 
On the average, irreversibility is suppressed quantum-mechanically compared with the semiclassical expectation.
But more interestingly, the dependence of $N_{\text{states}}$ on $\dot{\Phi}$ becomes 
erratic, indicating a crossover to a regime of chaos-assisted-depletion.  
This crossover is further reflected in the timing of the depletion, 
as we already saw in \Fig{FigTrimerDepletionTimeVsRate}.

\rmrk{\section{Common approximations that exclude `chaos'}}

\subsection{Two orbital approximation}

As we sweep the parameter $\Phi$, orbitals $k_0$ and $k_{+}$ cross each other. 
It is therefore natural to adopt TOA as in \cite{SwallowBrand}. 
This naturally leads to an effective 2 sites (dimer) problem as in \cite{cst},
that can be regarded as second-quantized version of the well known nonlinear LZ problem \cite{LZ-Niu-PRL,LZ-Niu-PRA}. 

With TOA, the 3rd term in \Eq{eHb} does not generate transitions between orbitals. 
Therefore we need a tilt $\epsilon{\ne}0$ in order to get non-trivial dynamics. 
Indeed this was the approach in \cite{SwallowBrand}. 
But clearly for a BHH ring we should have non-trivial dynamics 
even without a tilt. So clearly TOA is an over-simplification. 
Nevertheless one may wonder whether with $\epsilon{\ne}0$ 
there is a regime such that TOA makes sense.
We address this secondary question in \App{ApG}.

\begin{figure*}
\centering
\includegraphics[width=5.5cm]{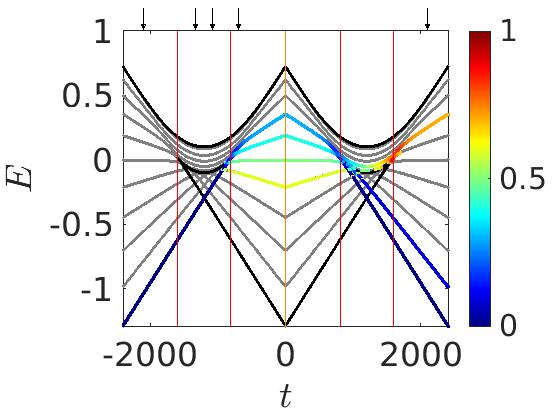}
\includegraphics[width=5.5cm]{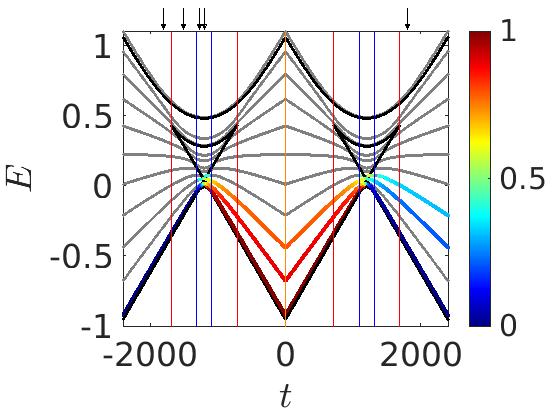}
\includegraphics[width=5.5cm]{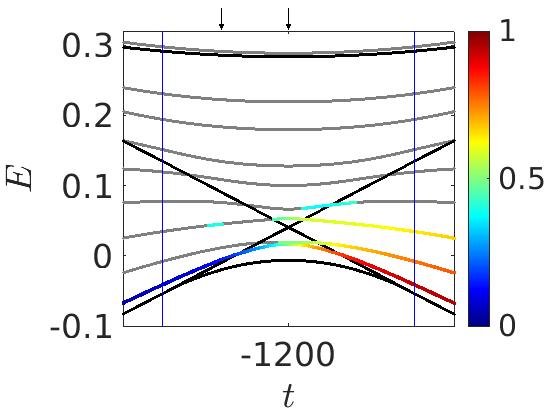} \\
\includegraphics[height=3cm]{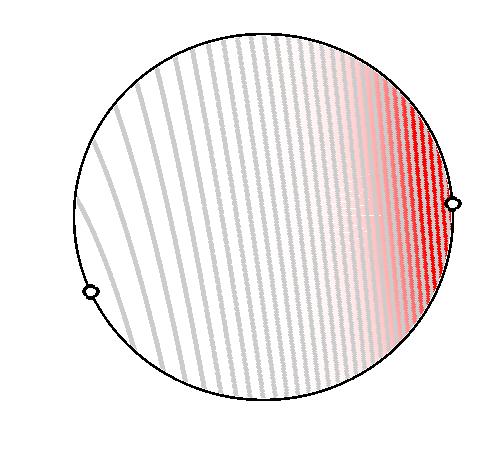}
\includegraphics[height=3cm]{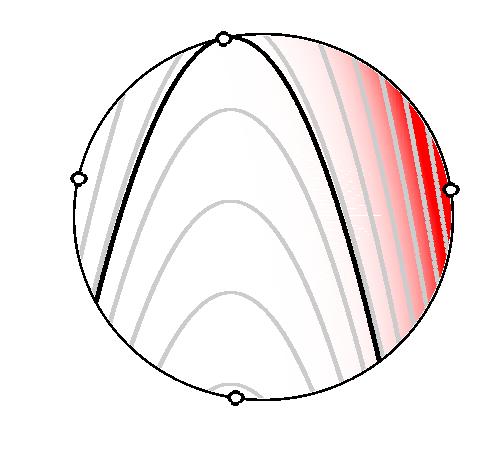}
\includegraphics[height=3cm]{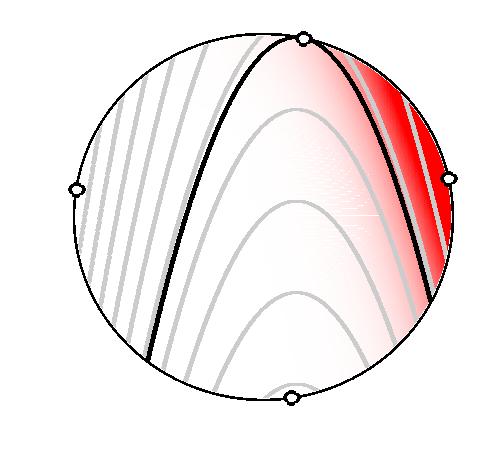}
\includegraphics[height=3cm]{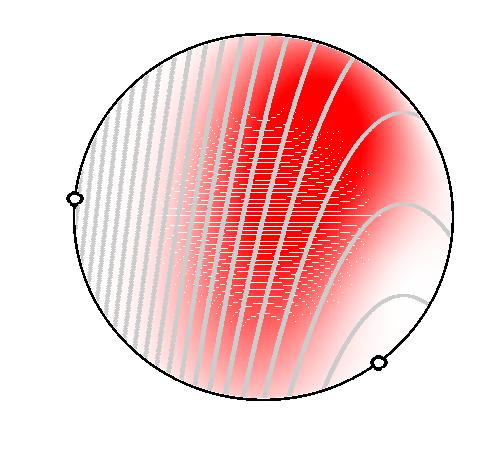}
\includegraphics[height=3cm]{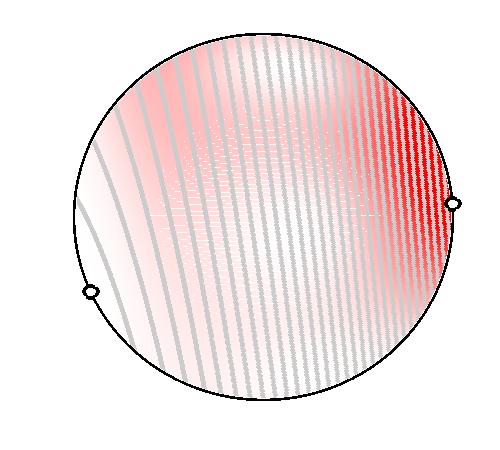}
\includegraphics[height=3cm]{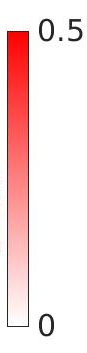} \\
\includegraphics[height=3cm]{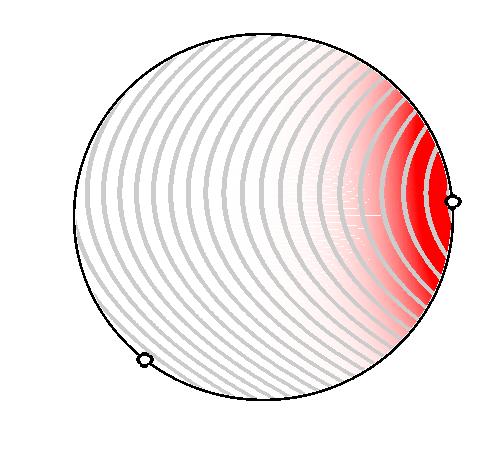}
\includegraphics[height=3cm]{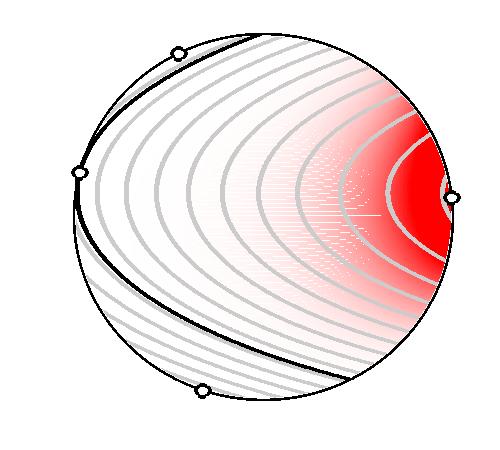}
\includegraphics[height=3cm]{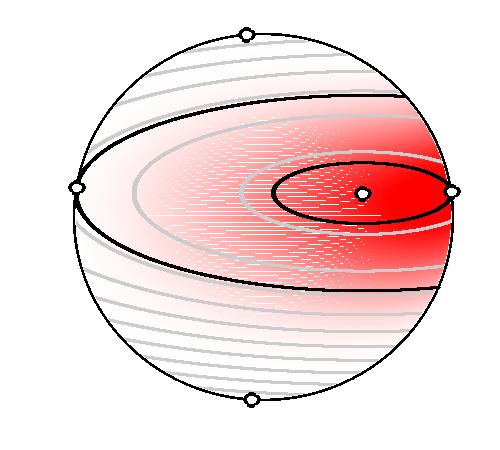}
\includegraphics[height=3cm]{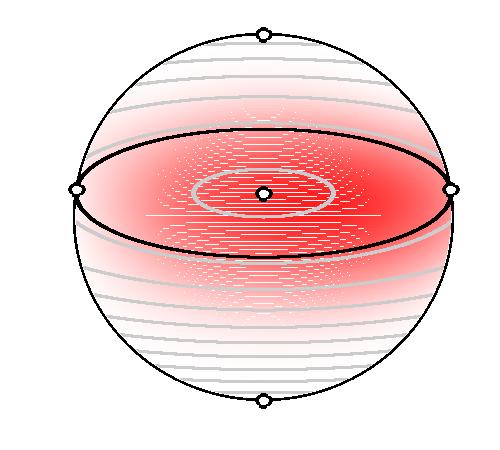} 
\includegraphics[height=3cm]{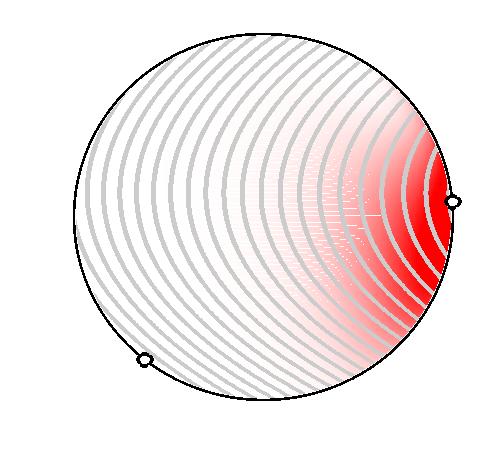}
\includegraphics[height=3cm]{naive_colorbar}
\caption{ \label{FigDimerEnergyVsTimeQM} \label{FigDimerPhasespace} 
{\bf Quantum simulations for the dimer.} 
We consider $N{=}10$ particles whose dynamics is generated by the Hamiltonian \Eq{eHdimer}.
The upper panels are for diabatic ejection scenario (left), 
and relay shuttling (middle), and zoom of the latter (right). 
The units of time are such that $K{=}1$, 
and the \rmrk{tilt} is $\epsilon{=}0.2$.  
The interaction parameters are given respectively 
by \Eq{Utoa} with ${u{=}3.45}$ and by \Eq{Ubogo} with ${u{=}2.3}$, 
with $L{=}3$. 
The sweep is from $\mathcal{E}{=}2$ to  $\mathcal{E}{=}-2$ and back to $\mathcal{E}{=}2$, 
with rate  $\dot{\mathcal{E}}=1/600$.  
Energy levels $E_{\nu}$ are plotted versus time. 
Levels whose  $p_{\nu}$ is vanishingly small are in gray.
The other levels are color-coded by~$\braket{n}_{\nu}$.
The energies of the minima, maxima and seperatrices are indicated by black lines.
Bifurcation points are indicate by vertical lines.  
Snapshots of the evolution are taken at times that are indicated by small black arrows, 
and placed at the 2nd row (diabatic ejection) and at the 3rd row (relay shuttling). 
At each snapshot we plot the Husimi representation \Eq{eHus} (red is high intensity) 
of the quantum state, using $(S_x,S_z)$ phase-space coordinates.  
We overplot energy contours of the Hamiltonian, 
and indicate in black the extremal points and the seperatrices.   
}
%
%
%
\centering
\includegraphics[width=8cm]{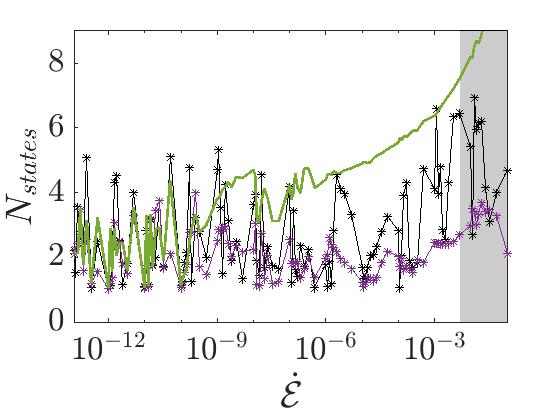}
\includegraphics[width=8cm]{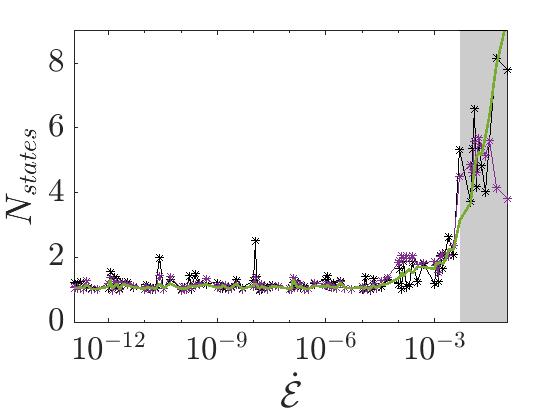}
\caption{ \label{FigDimerSpreadVsRate}
{\bf Irreversibility vs sweep-rate for the dimer.} 
For simulations as in \Fig{FigDimerEnergyVsTimeQM}, 
we plot $N_{\text{states}}$ versus the sweep rate $\dot{\mathcal{E}}$ 
at the end of the forward sweep (purple),  
and at the end of the reversed sweep (black). 
The left panel is for the diabatic ejection scenario, 
and the right panel is for the relay shuttling scenario. 
Additionally we plot (in green) the average level index at the end of the forward sweep. 
The non-quasistatic region (gray) is determined by inspection 
of the $N_{\text{orbitals}}$ plot of \App{ApN}.  
}
\end{figure*}

\subsection{Bogolyubov approximation}

The Bogolyubov approximation keeps in \Eq{eHb} transitions of {\em pairs} 
from the $k_0$ condensate to the $k_{\pm}$ orbitals. 
The textbook version further makes the substitution ${b_0 \mapsto \sqrt{N}}$, 
but we avoid below this over-simplification. 
Either way, it is clear that the Bogolyubov approximation 
implies that in the absence of tilt ($\epsilon{=}0$) 
the occupation imbalance ($M$) is a constant of motion. 
Consequently, for the ${L=3}$ trimer 
(or for any ring if we keep the 3 lowest orbitals ${k_0,k_{+},k_{-}}$) 
the BHH becomes formally identical to a generalized dimer Hamiltonian, 
that differs from the standard TOA dimer. 

We present the derivation of the effective $\mathcal{H}_{\text{dimer}}$ 
in \App{ApB}, and further discuss it below. 
The same $\mathcal{H}_{\text{dimer}}$ can be exploited to 
simulated the TOA dynamics using appropriate set of effective parameters, 
and to simulate the Bogolyubov dynamics using a different set of effective parameters. 
The dynamics that is generated in the two cases  
is illustrated in \Fig{FigDimerEnergyVsTimeQM}. 
One observes that the TOA dynamics (with tilt) features {\em diabatic ejection}. 
As opposed to that, the Bogolyubov-approximated dynamics features what we call {\em relay shuttling}.  
The snapshots of the evolution that are provided in \Fig{FigDimerEnergyVsTimeQM} 
correspond to the scenarios that have been caricatured in \Fig{FigBifurcations}. 

Coming back to \Fig{FigTrimerDepletionTimeVsRate} we observe  
that the $t_d$ of the  Bogolyubov (black) line 
agree with that of the blue line, 
but not with that of the red line.  
This implies that in the latter case (very slow sweep) 
the depletion mechanism is {\em not} a relay-shuttling process.

\subsection{The generalized dimer problem}

Both the TOA (with tilt) and the Bogolyubov approximation (with or without a tilt) 
lead to an effective dimer problem. See \App{ApB} and \App{ApG}. 
The dimer Hamiltonian can be written using generators of spin-rotations.
Namely, $S_z$ is defined as half the occupation difference in the site representation,  
while ${S_x=(n_0-n_1)/2 = (N/2)-n }$ is half the occupation difference
in the momentum orbital representation. Thus, $S_x$ is merely a shifted version 
of the depletion coordinate. What we call generalized dimer Hamiltonian 
contains two distinct interaction terms: 
\beq  \label{eHdimer}
\mathcal{H}_{\text{dimer}} 
=
- \mathcal{E} S_x
- \epsilon S_z
+ U_{\parallel} S_z^2 
+ U_{\perp} S_x^2 
\eeq
In \App{ApC} we show that the TOA reduces to this form with 
\beq \label{Utoa}
U_{\parallel} =0 
, \ \ \ \ \ \ \ 
U_{\perp} = -\frac{1}{L}U
, \ \ \ \ \ \ \ \text{[TOA]}
\eeq
In contrast, the Bogolyubov approximations features, due to the pairing interaction,
\beq \label{Ubogo}
U_{\parallel} = \frac{2}{L} U   
, \ \ \ \ \ \ \    
U_{\perp} = \frac{1}{4L} U 
, \ \ \ \ \ \ \ \text{[Bogolyubov]}
\eeq
The detuning parameter $\mathcal{E}$ reflects the 
excitation energy of the condensate. 
For the TOA it is ${\mathcal{E}=\mathcal{E}_{+} - \mathcal{E}_{0}}$, 
while for Bogolyubov it is 
\beq \label{eEPhi}
\mathcal{E}(\Phi) \ = \ \frac{1}{2}\left( \mathcal{E}_{+} + \mathcal{E}_{-} \right) - \mathcal{E}_{0}  + \frac{NU}{4L} 
\eeq
As $\Phi$ is increased, $\mathcal{E}$ decreases,  
and at $\Phi_{\text{swp}}$ it swaps sign, 
namely ${\mathcal{E}(\Phi_{\text{swp}} )=0}$.  
The swap location is indicated by green vertical line in the time axis of our figures.  

We further show in \App{ApF} that the bifurcation scenario 
depends on the relative magnitudes of the $U$-s.
The parameters ${U_{\perp} }$ and ${U_{\parallel} }$ have the same sign (the latter is zero for TOA).    
Accordingly, phase space contours on the Bloch spheres are ellipses (or parabolas) in the $(S_x,S_z)$ coordinates.
If we vary the control parameter $\mathcal{E}(\Phi)$, 
there are two different bifurcations scenarios 
depending which interaction is larger.  
The two scenarios are compared in \Fig{FigDimerEnergyVsTimeQM} and \Fig{FigDimerSpreadVsRate}, 
and further discussed below.

Consider the TOA, for which we have ${|U_{\perp}| > |U_{\parallel}| }$. 
For large $\mathcal{E}$ the lowest energy is in the East pole, 
which supports condensation in orbital \#0. 
As $\mathcal{E}$ is decreased, a bifurcation appears at the West hemisphere, 
with separatrix that move to the East. This leads eventually 
to a diabatic ejection of the condensed cloud.  
We show in \App{ApF} that the pertinent bifurcations happens at 
\beq \label{eEcDia} 
\mathcal{E}_c = \left[ \left(|U_{\parallel}-U_{\perp}| N\right)^{2/3} - \epsilon^{2/3} \right]^{3/2}
\eeq

Consider the Bogolyubov approximation,  
for which we have ${|U_{\perp}| < |U_{\parallel}| }$.
Here two bifurcations take place: 
The first bifurcation appears at the West hemisphere, 
and is formally the same as that of \Eq{eEcDia}.
The same expression for $\mathcal{E}_c$ applies. 
However, this bifurcation has no significance, 
as implied by \Fig{FigDimerEnergyVsTimeQM}.
It is followed by a second bifurcation of the East pole 
that for zero \rmrk{tilt} is determined by 
the condition ${ \mathcal{E}(\Phi) = \mathcal{E}_{\text{dyn}} }$,
where ${ \mathcal{E}_{\text{dyn}} = N U_{\perp} }$. 
For non-zero  \rmrk{tilt} we derive in \App{ApF} the more general expression  
\beq \label{eEcSht} 
\mathcal{E}_{\text{dyn}} \ = \ \left[ 1-\left(\frac{\epsilon}{U_{\parallel}N}\right)^2 \right]^{1/2} NU_{\perp}
\eeq 
This bifurcation signifies the loss of dynamical-stability 
of the condensate (elliptic fixed-point becomes hyperbolic), 
and therefore the above condition can be used to determine $\Phi_{\text{dyn}}$.     
Due to the bifurcation a new minimum is born, 
and a relay-shuttling process is initiated.  
Subsequently, at $\Phi_{\text{swp}}$, 
there is a swap of seperatrices, and consequently, 
hereafter, the minimum that had bifurcated from the East 
belongs to the basin of the West. 
The net effect is relay-shuttling from East to West 
that ends when ${ \mathcal{E}(\Phi)= -\mathcal{E}_{\text{dyn}} }$.
This scenario is illustrated in \Fig{FigDimerEnergyVsTimeQM}.

\begin{figure*}
\centering
(a) \hspace*{16cm}  \\ \vspace*{-4mm}
\includegraphics[width=7cm]{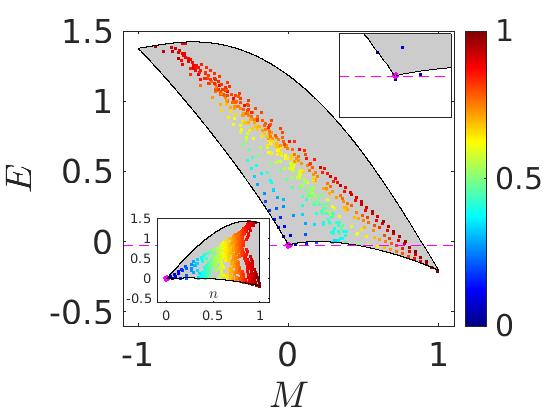}
\includegraphics[width=7cm]{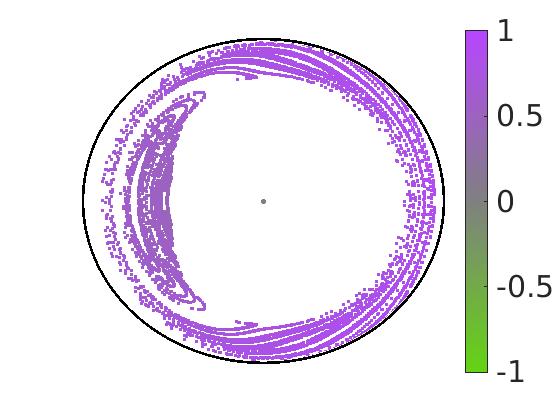} 
\\
(b) \hspace*{16cm}  \\ \vspace*{-4mm}
\includegraphics[width=7cm]{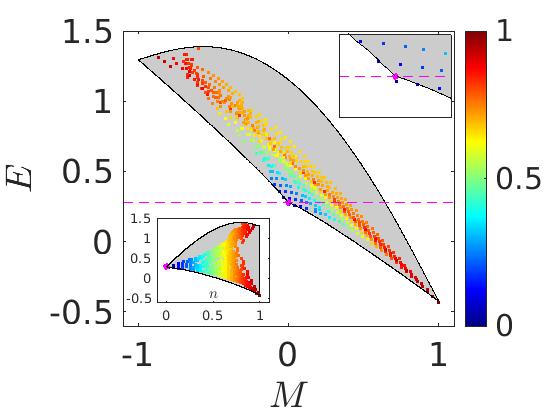}
\includegraphics[width=7cm]{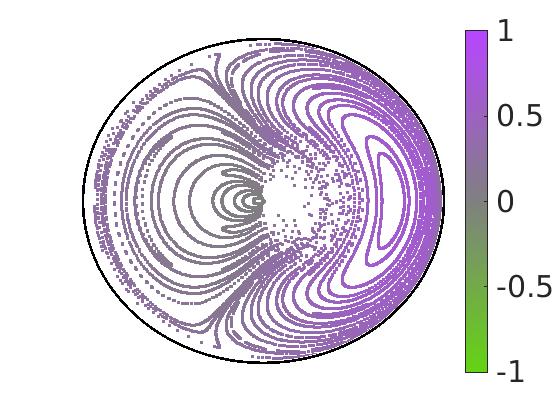} 
\\
(c) \hspace*{16cm}  \\ \vspace*{-4mm}
\includegraphics[width=7cm]{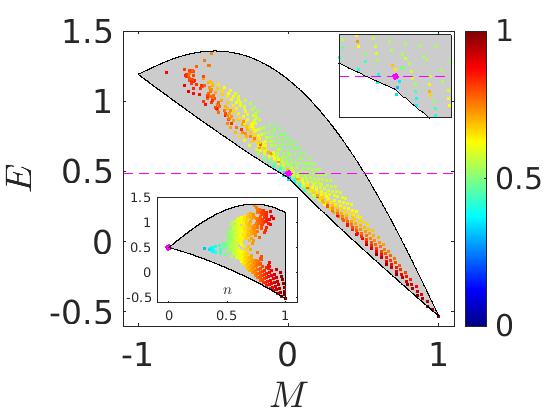}
\includegraphics[width=7cm]{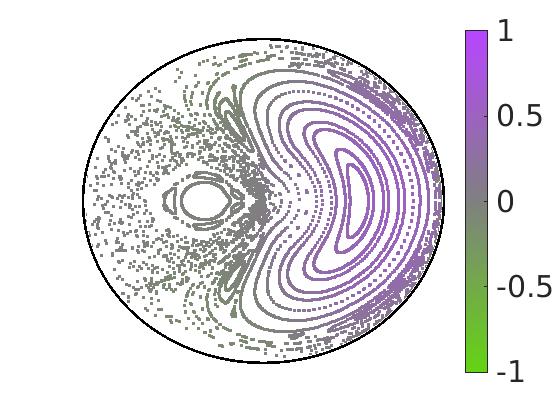} 
\caption{ 
{\bf Mixed chaotic phase space.}
The energy landscape of the BHH trimer for $u{=}2.3$. 
Rows (a), (b) and (c) are for ${\Phi = 1.1\pi,1.4\pi,1.6\pi}$.
Respectively they feature energetic stability, 
dynamical stability, and instability of the condensate.    
Each point in the left panels represents an eigenstate 
that is positioned according 
to its $E_{\nu}$ and $\braket{M}_{\nu}$, 
and color-coded according to its $\braket{n}_{\nu}$. 
Looking on the classical Hamiltonian ${\mathcal{H}(n,\varphi; M,\phi)}$, 
for each $M$ we find the floor (minimum) and the ceiling (maximum) of the energy, 
and get the Black solid lines that bound the spectrum from below and from above. 
The ${n{=}M{=}0}$ central fixed-point is indicted by a pink dot. 
Its vicinity is zoomed in the upper inset,  
and its energy is indicted by a dashed line.
For this energy a {\em Poincare section} of phase space is displayed in the right panel,  
where each section-point of a trajectory is color-coded by its~$M$, 
and displayed using its ${(n,\varphi)}$ as polar coordinates. 
In row (c) we have ${ \Phi > \Phi_{\text{dyn}} }$, 
for which the central point is an unstable saddle immersed in chaos.
Therefore it cannot support an eigenstate. 
This observation is better delivered by the lower inset, 
where the same spectrum is plotted with~$n$ serving as horizontal axis. 
} 
\label{fLandscape} 
\end{figure*}

\section{The manifestation of chaos}

Once we go beyond the Bogolyubov approximation, 
the imbalance $M$ is no longer a constant of motion.
Using action angle variables ($n$,$M$, and their conjugates)  
it is possible to express the 3-orbital Hamiltonian 
as the sum of integrable Bogolyubov term  
${ \mathcal{H}^{(0)}(n,\varphi; M) }$ that conserves $M$, 
and additional terms ${ \mathcal{H}^{(\pm)} }$ that spoil the integrability. 
See \cite{qtp} and \App{ApC} for explicit expressions.
The ${ \mathcal{H}^{(\pm)} }$ terms allow slow depletion of the cloud   
by drifting away from $M{=}0$.

Rows (a-c) of \Fig{fLandscape} clarify how phase-space changes as $\Phi$ is varied. 
It is the inspiration for the caricature in the right panel of \Fig{FigBifurcations}.
Snapshots are taken after  $\Phi_{\text{mts}}$, after $\Phi_{\text{stb}}$, 
and after $\Phi_{\text{dyn}}$. It shows how the $n{=}0$ fixed-point 
changes from metastable minimum to elliptic fixed-point and then becomes unstable.
We also have an indication for the emerging shuttling island.
The small island that we see in the right panel of row~(c) is in fact a section 
of torus that resides above the captured cloud. 
The latter can be located in a Poincare section at a slightly lower energy (not displayed).
The chaotic region allows an optional depletion process that we further discuss 
in the next paragraph. 

A necessary condition for chaos-assisted depletion is to have a potential floor 
that goes down from $n{=}M{=}0$ in the $M{\ne}0$ direction.  
This is the Landau criterion for instability of the superflow.  
Namely, $n{=}0$ becomes a saddle rather than a local minimum in the energy landscape.   
The Landau-instability is encountered once we cross $\Phi_{\text{stb}}$, 
which is indicated by the blue vertical line in the time axis of our figures.  
Bogolyubov analysis \cite{qtp} provides the explicit expression  
\beq \label{PhiSTB}
\Phi_{\text{stb}} = \ \ 3 \arccos{\left(\frac{1}{6} \left(\sqrt{u^2+9}-u\right)\right)}
\eeq  
where $u=NU/K$ is the dimensionless interaction strength.  
But we have to remember that only later, at $\Phi_{\text{dyn}}$, 
the $n{=}0$ location becomes dynamically unstable, as shown in \Fig{fLandscape}c. 
This means that for ${ \Phi_{\text{stb}} <  \Phi < \Phi_{\text{dyn}} }$
only the outer piece of the cloud can drift away from $M{=}0$ via the chaotic region. 
The implied branching is clearly demonstrated in \Fig{FigTrimerEnergyVsTimeSC} 
and optionally in \Fig{FigTrimerDynamicsSplitting} of \App{ApBranching}.  

The splitting of cloud, 
into an $M{=}0$ shuttling branch and $M{\ne0}$ chaotic spreading, 
is responsible for the crossover to chaos-assisted depletion. 
The latter is a very slow process, 
and therefore becomes noticeable 
only for very slow sweep rate.    
It is clearly distinct from shuttling, 
because it starts earlier, at $\Phi_{\text{stb}}$, 
unlike the shuttling that starts at $\Phi_{\text{dyn}}$. 

In the reversed sweep we see once again this branching effect. 
In fact it is more conspicuous on the way back: 
the cloud stretches further in the $M$ direction, 
which becomes possible because the ceiling of the potential 
is going up, hence not blocking further expansion.
An optional way to illustrate this branching  
is provided by \Fig{FigTrimerDynamicsSplitting} of \App{ApBranching}.

\section{Mechanisms for irreversibility}

In linear response theory (Kubo formalism), irreversibility is related to accumulated deviation from adiabaticity. It is controlled by the ratio between the sweep rate and the natural frequency of the driven system. This picture assumes that the cloud follows an evolving adiabatic-manifold in phase-space.    
In the quasistatic limit, linear response theory implies reversibility. But this picture breaks down if during the sweep a violent event takes place. In the nonlinear LZ problem the local minimum is diminished at a particular moment of the sweep process due to an inverse saddle-node bifurcation, see \Fig{FigBifurcations}, consequently the cloud is ejected and stretched along the fading seperatrix. This is what we call {\em diabatic ejection}. On the way back the cloud can split between two regions as implied by the Kruskal-Neishtadt-Henrard theorem  \cite{Kruskal,Neishtadt1,Timofeev,Henrard,Tennyson,Hannay,Cary,Neishtadt2,Elskens,Anglin,Neishtadt3}. This type of dynamics is reflected in the quantum dynamics, 
see \Fig{FigDimerEnergyVsTimeQM} for demonstration.       

In the problem under consideration, {\em diabatic ejection} is an artefact of the TOA. Instead we find that the Bogolyubov approximation predicts {\em relay-shuttling}. A gentle type of irreversibility can arise when the shuttling process starts or ends (pitchfork bifurcations). See \Fig{FigDimerSpreadVsTime-App} of \App{ApN} for demonstration. A quantitative comparison of the irreversibility that is associated with the two mechanisms is provided in \Fig{FigDimerSpreadVsRate}.
  
As we already discussed, for very slow sweep a different depletion mechanism takes over, that goes beyond Bogolyubov, namely, chaos-assisted depletion. This mechanism gives rise to ``free expansion" of the cloud in phase space ($M$ is not constant of motion). Furthermore, once the sweep is reversed the cloud undergoes a conspicuous branching process, as discussed previously for \Fig{FigTrimerEnergyVsTimeSC} and \Fig{FigTrimerDynamicsSplitting} of \App{ApBranching}. Thus,  irreversibility is extremely enhanced in the semiclassical simulations. Quantitatively this has a modest manifestation in the quantum mechanical case. On the other hand, we observe a novel regime of quantum irreversibility that  exhibits ``quantum chaos" characteristics and breakdown of QCC that we further discuss below.

\section{Universal Quantum Fluctuations}

Classical evolution of expectation values reflect ergodization.
Namely, fluctuations are completely smoothed away if we wait enough time.
As opposed to that, quantum fluctuations persist and are not smoothed away.
This means that quantum mechanically the quasistatic limit does not exist. 
At any moment the state of the system cannot be regarded as stationary.   
In \Fig{FigTrimerSpreadVsWaiting-App} of \App{ApN}  
we demonstrate the dependence of $N_{\text{states}}$ on the waiting time~$T$.  
The same fluctuations are reflected if we plot $N_{\text{states}}$ versus $\dot{\Phi}$. 
We re-emphasize that such fluctuations are absent in the semiclassical simulations. 
(Therefore we set $T{=}0$ in the semiclassical simulations of \Fig{FigTrimerEnergyVsTimeSC}.)

\section{QCC and its breakdown}

We already pointed out that the semiclassical dynamics is reflected in the 
quantum simulations, see \Fig{FigTrimerEnergyVsTimeQM}. 
The term ``reflected" does not imply ``correspondence". 
We would like to explain the observed breakdown of QCC for slow sweep. 

For an extremely slow sweep (that cannot be realized in practice), 
the quantum dynamics would follow the ground state. 
This can be regarded as a  {\em quantum detour} 
of the classical non-adiabatic arena that was looming ahead. 
For realistic sweep rate the dynamics follows {\em diabatically} 
the metastable minimum. But still the probability can leak 
to levels that are crossed along the way.
This early leakage becomes more probable as the forbidden-area shrinks (low energetic barrier), 
and definitely once it is replaced by dynamical barriers of the Kolmogorov-Arnold-Moser (KAM) type \cite{tunnleKAM,tunnleHeller}.

The lifetime~$\tau$ of the condensate can be extracted 
from the local density of states (LDOS) of the Hamiltonian, see \App{ApH}. 
The interesting range, as explained above, is ${ \Phi_{\text{stb}} < \Phi <  \Phi_{\text{dyn}} }$.
In this range the classical cloud has a piece that is trapped 
on a dynamically stable island, and therefore cannot decay.  
But quantum mechanically the cloud can tunnel through the KAM barriers, 
and therefore has a finite lifetime~$\tau(\Phi)$. 

We are now equipped to estimate the border between the various $\dot{\Phi}$ regimes. 
The quantum adiabatic regime is determined by the standard condition ${|d\mathcal{H}/dt|<\kappa^2}$, 
where $\kappa$ is the tunnel coupling, that determines the level splitting. 
As discussed earlier this condition is never satisfied in practice due to the smallness of~$\kappa$. 
Using ${\alpha \equiv |d\mathcal{H}/d\Phi| \sim K }$,  
we can re-write the adiabatic condition as follows, 
\beq \label{eTau}
\tau(\Phi)  \ <  \ \frac{\Delta\Phi}{\dot{\Phi}} 
\eeq
where ${ \Delta\Phi = \kappa/\alpha }$, 
is the parametric width of the avoided crossing, 
and ${\tau \sim 1/\kappa }$ is the time to make a Rabi transition.  
We can extend this reasoning to the Fermi-Golden-Rule (FGR) regime 
where $\kappa$ becomes larger than the effective levels spacing $\Delta_0$. 
The latter refers to the participating levels of the LDOS. 
There we expect ${\tau = 1/\gamma}$, 
with ${\gamma=2\pi \kappa^2/\Delta_0}$. 
The condition for having an escape before $\Phi_{\text{dyn}}$ 
is obtained from \Eq{eTau}, 
and implies a crossover at ${\dot{\Phi} \sim 10^{-4} \pi }$,  
in rough agreement with \Fig{FigTrimerDepletionTimeVsRate}.

\section{Discussion} 

Considering a closed classical Hamiltonian driven system, such as a particle in a box with moving wall (aka the piston paradigm), the common claim in Statistical Mechanics textbooks is that quasistatic processes are adiabatic, with vanishing dissipation, and hence reversible. This statement is indeed established for {\em integrable} \cite{Landau} and for fully {\em chaotic} systems \cite{Ott1,Ott2,Ott3,Wilkinson1,Wilkinson2,crs,frc}. But generic systems are neither integrable nor completely chaotic. Rather they have {\em mixed phase space}. For such system the quasistatic limit is not adiabatic \cite{Kedar1,Kedar2,apc,lbt,qtp}, and therefore we expect irreversibility. This irreversibility can be regarded as the higher-dimensional version of separatrix crossing \cite{Kruskal,Neishtadt1,Timofeev,Henrard,Tennyson,Hannay,Cary,Neishtadt2,Elskens,Anglin,Neishtadt3}, where the so-called Kruskal-Neishtadt-Henrard theorem is followed.

In the present work we wanted not just to expand the analysis of classical irreversibility, but also to explore the quantized version. We asked whether the distinct mechanisms of classical irreversibility are reflected in the quantum mechanical arena, and how this reconciles with the observation that quantum dynamics, unlike classical dynamics, is always reversible in the strict quasistatic (adiabatic) limit. 
Our main observations are as follows:  
{\bf (1)}~The TOA, and the associate LZ picture, do not provide a proper framework for the analysis of the depletion process. 
We need at least 3 orbitals in order to capture the essential features of the dynamics. 
This means that we are dealing here with a ``quantum chaos" problem.    
{\bf (2)}~The Bogolyubov approximation, unlike the naive TOA, implies gentle type of irreversibility 
that is related to relay shuttling, and not to diabatic ejection.  
{\bf (3)}~Beyond the Bogolyubov approximation we have chaos-assisted mechanism that competes 
with the relay shuttling process. This mechanism becomes dominant in the deeper quasistatic regime. 
{\bf (4)}~Accordingly, with regard to the sweep rate, one has to distinguish between 
non-quasistatic regime; relay-shuttling regime; chaos-assisted regime; and quantum adiabatic regime.
For a manybody condenstate, the latter is not accessible in practice. 
{\bf (5)}~Quantum features dominate the quantum adiabatic regime and the chaos-assisted regime. 
The most prominent effect can be described as a version of universal quantum fluctuations.        
{\bf (6)}~In the same regime, breakdown of QCC is conspicuous.  
It is related to leakage of probability along the diabatic transitions.    
Such leakage does not exist in the semiclassical simulations.

On the practical side one observes that the optimization of a protocol is related to the crossovers between the various regimes. Sweeping a control parameter `too fast' takes us out of the quasistatic regime, while `too slow' is affected by chaos. UQF possibly can be exploited for fine-tuning, whose purpose is to minimized chaos-related irreversibility. In analogy with the claim that diagonalization of the Hamiltonian can provide in ``one shot" phase-space tomography \cite{csf}, also here we can say that relatively cheap quantum simulations, can provide information on the classical dynamics for a cloud of trajectories.

\appendix

\section{The BHH for the dimer}
\label{ApB}

The BHH \Eq{eHb} for an $L{=}2$ dimer is 
\beq
\mathcal{H}_{\text{dimer}} = && \sum_{j=0,1} \left[ \epsilon_j a^{\dag}_j a_j 
+ \frac{U}{2} a^{\dag}_ja^{\dag}_ja_ja_j  \right] 
\\
&& -\frac{K}{2} \left( a^{\dag}_1a_0 + a^{\dag}_0a_1 \right)
\eeq
In momentum representation (ground state orbital labeled as "0", 
and excited orbital labeled as "+") it takes the form    
\beq \label{eUdimer}
&& \mathcal{H}_{\text{dimer}} \ \  = \ \sum_{k=0,+}  \mathcal{E}_k n_k 
- \frac{\epsilon}{2} \left(b^{\dag}_0b_{+}+b^{\dag}_{+}b_0 \right)
\nonumber \\
&& + \frac{U}{4}(N-1)N + \frac{U_o}{2}n_{+}n_{0} + \frac{U_{\parallel}}{4} \left(b^{\dag}_{+}b^{\dag}_{+} b_{0}  b_{0} + \text{h.c.} \right) 
\ \ \ \ \ \ \ \ 
\eeq
with ${ \mathcal{E}_k = \mp (K/2) }$
and ${U_0{=}U_{\parallel}}{=}U$. 
Making the substitution ${b_j\mapsto \sqrt{n_j} e^{i\varphi_j} }$, 
and ${n_0=N{-}n}$, and ${n_{+}=n}$, we get   
\beq
\mathcal{H}_{\text{dimer}} &=& 
E_0 + \mathcal{E} n 
\ - \epsilon \sqrt{(N{-}n)n}\cos(\varphi)
\nonumber \\
&& +\frac{U_o}{2}(N{-}n)n + \frac{U_{\parallel}}{2} (N{-}n)n \cos(2\varphi) 
\ \ \ \ \ \ \ \ \ 
\eeq 
where $E_0=N\mathcal{E}_0 + (U/4)(N{-}1)N$ is a constant that can be dropped, 
and $\mathcal{E}=\mathcal{E}_{+}-\mathcal{E}_0 = K$ is the detuning of the two orbitals.

One can write the interaction term of \Eq{eUdimer} 
using generators of spin-rotations.  
We define ${S_x = (n_{+}{-}n_0)/2}$, 
while ${n_{+}{+}n_0=N}$, and use the identity 
\beq
S_{\Delta}^2 \ \equiv \ S_z^2-S_y^2 \ = \ \frac{1}{2} \left(b^{\dag}_{+} b^{\dag}_{+} b_0  b_0 + \text{h.c.} \right)
\eeq     
Dropping a constant we get
\beq 
\mathcal{H}_{\text{dimer}} =  
- \mathcal{E} S_x
\ - \epsilon S_z
\ - \frac{U_o}{2} S_x^2 
\ + \frac{U_{\parallel}}{2} S_{\Delta}^2 
\ \ \ \ \ 
\eeq
We substitute ${S_{\Delta}^2 = S_z^2-S_y^2}$, 
and in order to get rid of $S_y^2$   
exploit that ${S_x^2+S_y^2+S_z^2}$ is a constant of motion. 
Thus the final expression can be 
written as in \Eq{eHdimer}, 
with ${U_{\perp} = (U_{\parallel}{-}U_o)/2 }$.

\section{BHH interaction term for a ring}
\label{ApC}

The momentum index $k$ can be defined mod($L$) 
such that $k := (2\pi/L)k$ is the quasi momentum in standard units.
For a trimer this index takes the values ${k=0,\pm 1}$ or shortly ${k=0,\pm}$.  
The prime in the $k$ summation of \Eq{eHb}
implies that conservation of total momentum is required.
The interaction term can be arranged as follows:
\beq
&& \frac{1}{2}\sum_{k}^{'} b^{\dag}_{k_1}b^{\dag}_{k_2}b_{k_3}b_{k_4} 
\ = \ \frac{(N{-}1)N }{2}  
\ + \sum_{\braket{k,k'}} n_k n_{k'}  
\ \ \ \ \ \ \
\nonumber \\
&& 
\ \ \ \ \ \ \
\ + \ \mathcal{H}^{\text{pairing}} \ + \ \mathcal{H}^{\text{scattering}} 
\ \ \ \ \ \ \ 
\eeq
The second term reflects the cost of fragmentation. 
The summation is over pairs (without double counting).  
The last term includes scattering events that involve 4 different orbitals, 
while the pairing events involve only 3 orbitals (two $k$ particles split into $k \pm q$ orbitals, and vice versa).    
In the special case $L{=}3$, the scattering events are absent. 
Dropping a constant, we are left with  
\beq
\frac{U}{L} \left[ 
\sum_{\braket{k,k'}} n_kn_{k'}
+\sum_{k=0,\pm} \left( b^{\dag}_{k+1}b^{\dag}_{k-1}b_{k}b_{k}  + \text{h.c.} \right) 
\right]
\ \ \ \ 
\eeq
The Bogolyubov approximation is obtained 
if we keep in the second term only the $k{=}0$ transitions.
Then we get 
\beq \label{eUtrimer}
\frac{U}{L} \left[ n_0n_{+} + n_0n_{-} + n_{+}n_{-} + n_0 \sqrt{n_{+}n_{-}} \cdot 2\cos(2\varphi) \right] 
\ \ \ \ \ 
\eeq
Using $n$ and $M$ as coordinates this expression takes the form 
\beq 
\frac{U}{L} \left[ (N{-}n)n + \frac{1}{4}(n^2-M^2)  \right] 
\nonumber \\ 
+ \frac{U}{L} (N{-}n) \sqrt{n^2-M^2} \cos(2\varphi) 
\eeq
Setting $M{=}0$, the above terms are formally the same 
as that of the dimer, provided we allow different 
coefficients ${U_{o}=(3/(2L))U}$  and  ${U_{\parallel}=(2/L)U}$, 
and include the term $NU/(4L)$ in \Eq{eEPhi}.   
On the other hand, if we keep in \Eq{eUtrimer} 
only the $n_0$ and the $n_{+}$, 
we get the TOA where ${U_{o}=(2/L)U}$  and  ${U_{\parallel}=0}$.  

For completeness we write the full Hamiltonian, 
without tilt as $\mathcal{H}^{(0)} + \mathcal{H}^{(+)} + \mathcal{H}^{(-)} + \const$. 
The integrable part can be written as follows:  
\beq 
&& \mathcal{H}^{(0)} \ = \ - \frac{U}{12}M^2  + \mathcal{E}_{\perp} M + \mathcal{E}n 
\nonumber \\
&& + \frac{U_o}{2}\left(N{-}n\right) n +  \frac{U_{\parallel}}{2} \left(N{-}n\right) \sqrt{n^2{-}M^2} \cos(2\varphi) 
\hspace{13mm}
\eeq
where the detuning parameters are
\beq
\mathcal{E} &=& \frac{1}{2}\left( \mathcal{E}_{+} + \mathcal{E}_{-} \right) - \mathcal{E}_{0}  + \frac{NU}{4L} \\
\mathcal{E}_{\perp} &=& \frac{1}{2}\left( \mathcal{E}_{+} - \mathcal{E}_{-} \right) 
\eeq
The Non-Bogolyubov terms in the $L{=}3$ trimer Hamiltonian 
arise from interaction that involves pairs 
that move in or out of excited orbitals.  
In action angle coordinates the explicit expression for them is
\beq
\mathcal{H}^{(\pm)} = \frac{U}{3\sqrt{2}}\sqrt{(N{-}n)(n{\pm}M)}(n{\mp}M)\cos{\left(3\phi{\mp}\varphi\right)} 
\ \ \ \ \ \ 
\eeq 
The non-Bogolyubov terms spoil the integrability of the BHH, 
and generate chaotic motion in phase-space.

\begin{figure*}
\centering
\includegraphics[width=6cm]{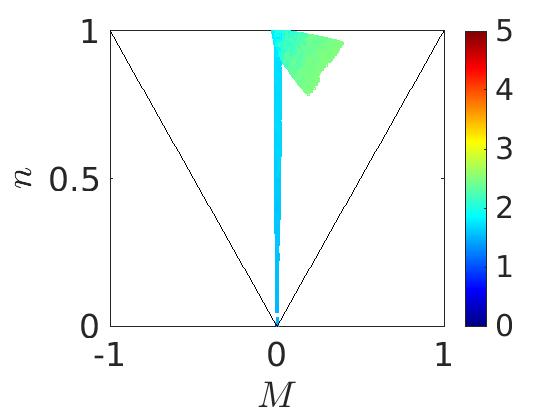} 
\includegraphics[width=6cm]{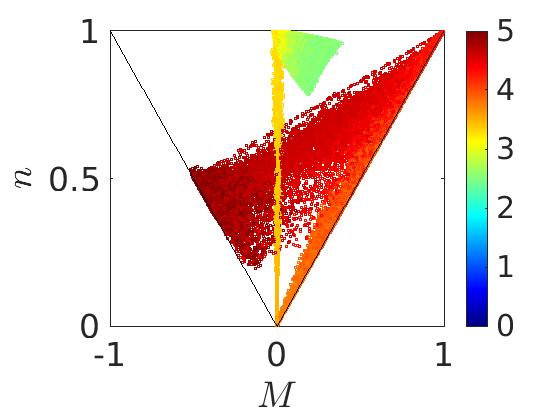} 
\\
\includegraphics[width=6cm]{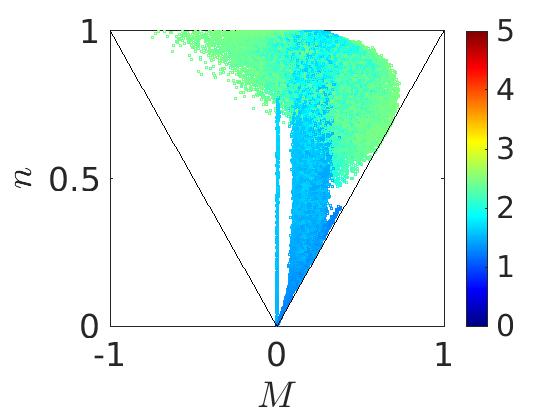}
\includegraphics[width=6cm]{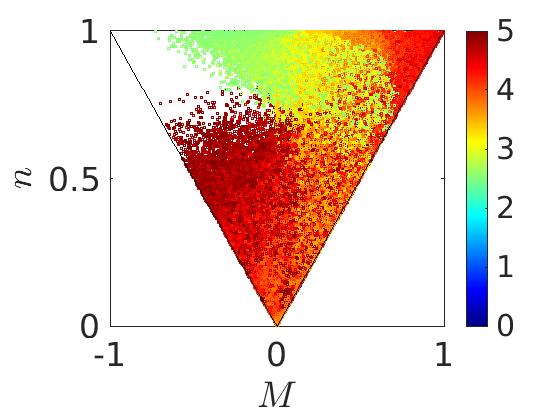}

\caption{ 
{\bf Evolution of the cloud in occupation space.} 
Optional plots for the semiclassical simulations of \Fig{FigTrimerEnergyVsTimeSC}.   
The left and the right panels are for the forward and for the reversed sweep,
with the optimal sweep rate $\dot{\Phi}{=}5\pi \cdot 10^{-4}$ (upper panels), 
and the very slow (non-optimal) forward sweep with $\dot{\Phi}=5\pi \cdot 10^{-5}$ (lower panels). 
\rmrk{The optimal sweep rate has been determined by the minimum of the black curve in \Fig{FigTrimerSpreadVsRate}b, 
meaning that it is slow, but not too-slow, such that relay-shuttling is still dominant.}    
\rmrk{The color-code reflects the time (the initial $t{=}0$ cloud is blue, the final cloud is red).}  
}
\label{FigTrimerDynamicsSplitting}
\end{figure*}

\section{Indicators of quantum chaos}
\label{AppSgntr}

The simplest indicator for ``quantum chaos" is level repulsion. 
In practice it is useful to define 
\beq
r_n \ = \ \frac{\min(\Delta_{\nu},\Delta_{\nu+1})}{\max(\Delta_{\nu},\Delta_{\nu+1})}   
\eeq
where $\Delta_{\nu}= E_{\nu+1}-E_{\nu}$ is the level spacing. 
The average of $r_n$ within an energy window is expected to be 
${\braket{r}\approx 0.536}$ for Wigner-Dyson (chaotic) statistics, 
as opposed to ${\braket{r}\approx 0.386}$ for Poissonian (nonergodic) statistics. 
A possibly better indicator is extracted from the matrix elements of the current operator
${ I_{\nu,\mu} = \BraKet{\nu}{(-\partial \mathcal{H}/ \partial \Phi)}{\mu} }$. 
The band profile of this matrix is related by Fourier transform to the current-current correlation function, 
and therefore its area ${s_{\nu} = \sum_{\mu (\ne \nu)}  |I_{\nu,\mu}|^2 }$ reflects the correlation time. Related measure, 
and discussion of its $L$ dependence, can be found in \cite{sgntr}.

\section{The branching of the cloud}
\label{ApBranching}

\Fig{FigTrimerDynamicsSplitting} shows how the evolution of the cloud of \Fig{FigTrimerEnergyVsTimeSC} 
look like in occupation space, using ${n,M}$ coordinates. This figure provides an optional view 
of the baranching: one piece of the cloud drifts away from $M{=}0$ starting at $\Phi _{\text{stb}}$, 
and another piece is shuttles along $M{=}0$ starting at $\Phi _{\text{dyn}}$. 
The branching is visible only for very slow sweep. In the forward sweep the drift stops 
after a short duration because the ceiling of the potential is going down,
hence blocking further expansion. But in the reversed sweep the ceiling of the potential is going up,    
and therefore the branching becomes conspicuous.  
\\

\begin{figure*}
\centering
\includegraphics[width=5cm]{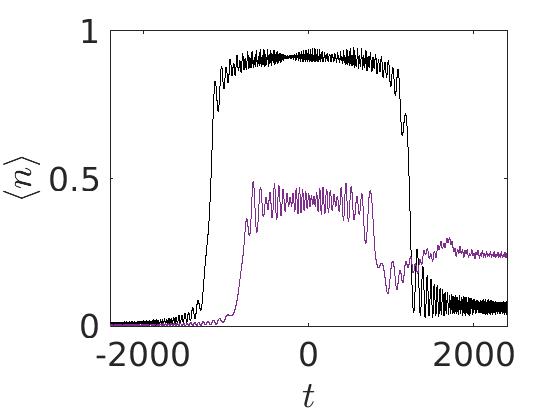}
\includegraphics[width=5cm]{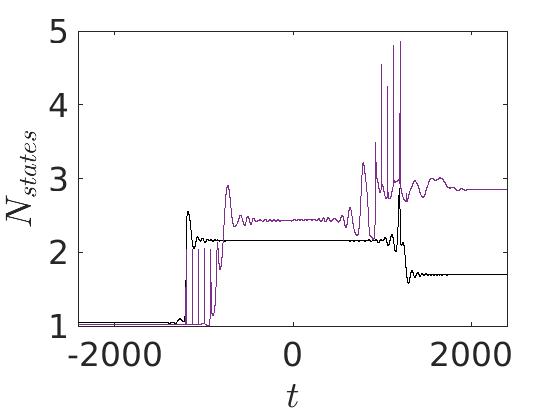}
\includegraphics[width=5cm]{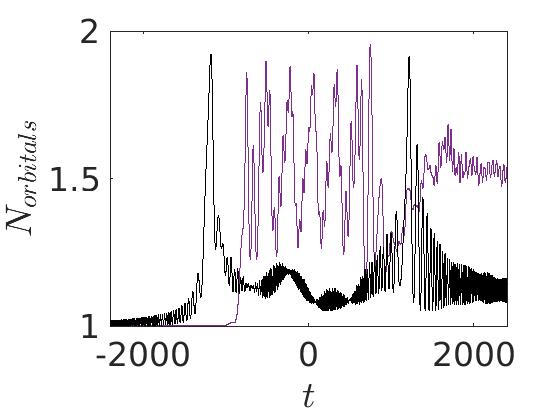}
\caption{ 
{\bf Depletion vs time for the dimer.} 
The depletion $\braket{n}$, and $N_{\text{states}}$ and $N_{\text{orbitals}}$, 
are plotted as a function of time 
for diabatic ejection scenario (purple) and for relay shuttling (black). 
Simulations parameters are as in \Fig{FigDimerEnergyVsTimeQM}. 
There is no relation between the two scenarios:  
they are combined in one plot for presentation purpose. 
The only meaningful comparison concerns the question 
whether the reversed sweep is capable of restoring the 
initial state.     
}
\label{FigDimerSpreadVsTime-App} 
%
%
\centering 
\includegraphics[width=5cm]{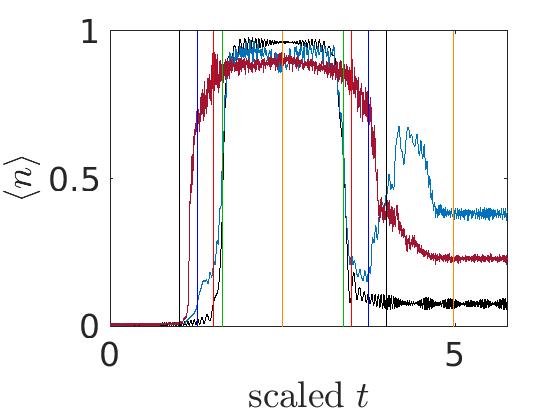}
\includegraphics[width=5cm]{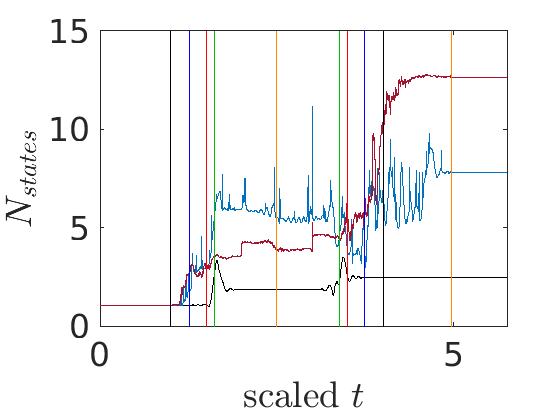}
\includegraphics[width=5cm]{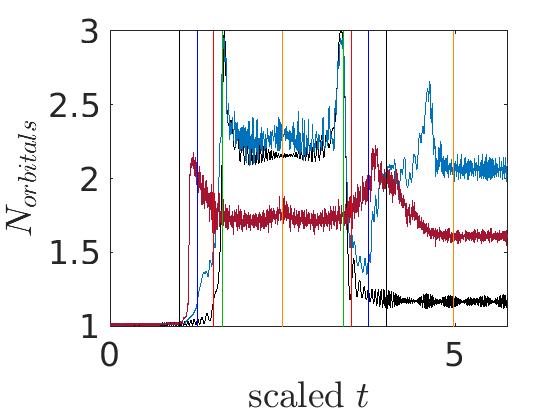}
\caption{ 
{\bf Depletion vs time for the trimer.} 
The depletion $\braket{n}$, and $N_{\text{states}}$ and $N_{\text{orbitals}}$, 
are plotted as a function of time 
for $\dot{\Phi}=5\pi \cdot 10^{-4}$ (blue), 
and for very slow rate $\dot{\Phi}=3.33\pi \cdot 10^{-7}$ (red).
The black line is generated with the Bogolyubov-approximated  
Hamiltonian $\mathcal{H}^{(0)}$ for $\dot{\Phi}=5\pi \cdot 10^{-4}$.
The other model parameters and the vertical lines are as in \Fig{FigTrimerEnergyVsTimeQM}.
}
\label{FigTrimerSpreadVsTime-App} 
\end{figure*}

\section{Participating orbitals}
\label{ApA}

The one-particle reduced probability matrix that is associated 
with a manybody state is 
${\rho_{k',k} =  (1/N)\braket{\bm{b}_{k}^{\dag} \bm{b}_{k'}} }$. 
We define 
\beq
N_{\text{orbitals}} \ =  \ \Big[ \tr(\rho^2) \Big]^{-1}  
\eeq
This is a measure for fragmentation.
For a manybody coherent state $N_{\text{orbitals}}{=}1$, 
meaning that all the particles are condensed in a single orbital.  
Semiclassically, such state can be pictured 
as a localized Gaussian-like distribution in phase-space. 
It is important to realize that the at the end 
of a relay shuttling process we get  $N_{\text{orbitals}}{=}1$ 
in the reduced dimer representation, 
but  $N_{\text{orbitals}}{=}2$ in the proper trimer representation, 
reflecting a Twin Fock state (half of the particles in each orbital).
At the swap we have $N_{\text{orbitals}}{=}3$.  
\App{ApN} provides plots of  $N_{\text{orbitals}}(t)$ 
and $N_{\text{states}}(t)$ for the protocols that are discussed in the main text.

For a dimer, the $N_{\text{orbitals}}{=}1$ coherent states 
are related as follow to the Fock states ${\ket{n}}$, 
\beq
\hspace*{-8mm}
|\theta,\varphi \rangle = \sum_{n=0}^N \sqrt{\left(\amatrix{N \\ n}\right)} 
\left[\cos{\frac{\theta}{2}}\right]^{N{-}n} 
\!\left[\sin{\frac{\theta}{2}}\right]^{n}   
e^{i n \varphi} \ |n \rangle 
\ \ \ 
\eeq
The Husimi function use this over-complete basis in order to represent 
the many-body quantum state on the ${(S_x,S_y,S_z)}$ Bloch sphere. 
Namely, it is defined as follows:  
\beq \label{eHus} 
Q(\theta,\varphi) = |\langle \theta,\varphi| \psi \rangle |^2
\eeq
If ${n = (N/2)-S_z}$ were the occupation-coordinate in the position (site) basis, 
then $\theta{=}0$ would be located at North pole. 
But we have defined ${n = (N/2) - S_x}$ 
as the occupations-coordinate in the momentum (orbital) basis. 
Therefore our ${n{=}0}$ is located at the East pole, 
which is re-defined as the origin for ${\theta}$.
Accordingly ${S_x=(N/2)\cos(\theta)}$.  
We plot images of the Husimi function using $(S_x,S_z)$ coordinates.
\\

\section{Depletion and spreading as a function of time}
\label{ApN} 

\begin{figure}
\centering
\includegraphics[width=6cm]{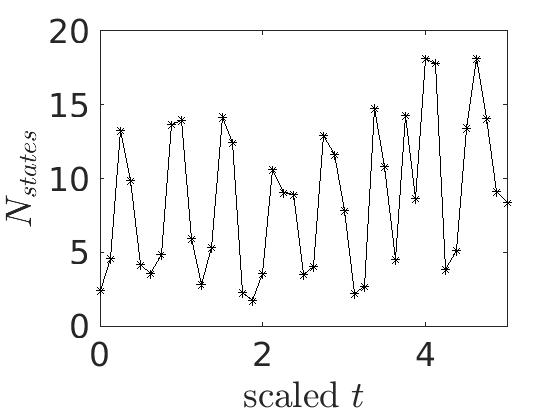}

\caption{ 
{\bf Irreversibility vs Waiting time.}
This is an additional panel for \Fig{FigTrimerSpreadVsRate}.
It illustrates the erratic dependence of $N_{\text{states}}$ 
on the waiting time for $\dot{\Phi}=5 \pi \cdot 10^{-6}$.   
In the main-text figure a few values of  $N_{\text{states}}$ 
are sampled for each $\dot{\Phi}$.  
}
\label{FigTrimerSpreadVsWaiting-App} 
\end{figure}

We present figures that provide examples for the temporal variation 
of $\braket{n}$ and $N_{\text{states}}$ and $N_{\text{orbitals}}$. 
\Fig{FigDimerSpreadVsTime-App}  is for the dimer simulations, 
while \Fig{FigTrimerSpreadVsTime-App}  and \Fig{FigTrimerSpreadVsWaiting-App} are for the trimer.
\Fig{FigDimerSpreadVsTime-App}  demonstrates that relay shuttling 
is rather reversible. As opposed to that, in diabatic ejection 
we have splitting in the revered sweep, which is reflected 
in $N_{\text{states}}$ and $N_{\text{orbitals}}$, 
and also spoils $\braket{n}$.   
In \Fig{FigTrimerSpreadVsTime-App}  we include a black line that is generated 
by the Bogolyubov-approximated Hamiltonian $\mathcal{H}^{(0)}$. 
This approximation if formally equivalent to the relay-shuttling
scenario of  \Fig{FigDimerSpreadVsTime-App}.
Note that its $t_d$ agree with the blue line, 
but not with the red line (very slow sweep), 
reflecting that different depletion scenarios are involved.
\\

\section{TOA vs Bogolyubov}
\label{ApG}

As far as $U$ is concerned, naive TOA for any ring (${L>2}$) gives no hopping. 
Therefore TOA implies that $\mathcal{H}$ of the rings takes the form of 
the dimer Hamiltonian \Eq{eUdimer} without the last term. 
We can compare it to the approximation that \cite{SwallowBrand} is using for a continuous ring
of length $2\pi R \equiv L a$.  To get this limit the lattice constant $a$ should be 
taken to zero, keeping $La$ constant.  
In this limit ${ K = (\mass a^2)^{-1} }$ is related to the mass of the particle. 
The gauge field is  ${\Phi = (\pi R^2) \times 2\mass \Omega }$,
where $\Omega$ is the rotation frequency.     
The single particle energies are 
\beq
\mathcal{E}_k = \frac{1}{2\mass R^2}\left(k - \mass \Omega R^2 \right)^2, 
\ \ \ \ \ \ k=\text{integer} \ \ \ \ \ 
\eeq
Hence, up to a constant, $\mathcal{E}$ is identified as the rotation frequency: 
\beq
\mathcal{E} \ \ = \ \  \mathcal{E}_1-\mathcal{E}_0   \ \ = \ \ \frac{1}{2 \mass R^2} - \Omega
\eeq

One wonders whether the discussion of 
``Nucleation in finite topological systems during continuous metastable quantum phase transitions" \cite{SwallowBrand} is flawed.  
In order to answer this question we have to appreciate 
the physical significance of the continuum limit ${L \rightarrow \infty }$ 
that was considered there.  
It is physically clear that ``rotation" of a flat clean ring  (that has neither tilt 
nor lattice potential) is an empty notion: nothing changes in the Hamiltonian. 
Furthermore, in this limit, chaos is not an issue (the ${L \rightarrow \infty }$ limit is integrable).
The physics that we discuss becomes relevant as $L$ becomes finite, 
and irreversibility is most pronounced for $L{=}3$.

\begin{figure}
\centering
\includegraphics[width=4cm]{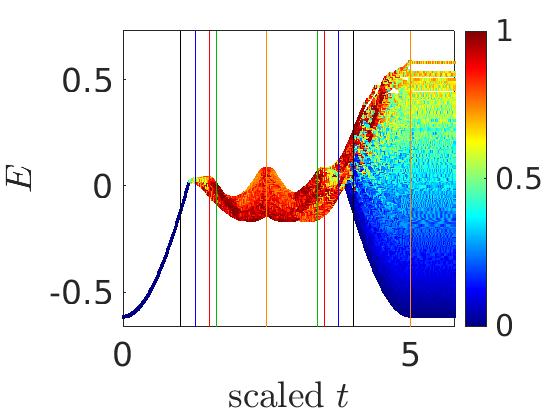} 
\includegraphics[width=4cm]{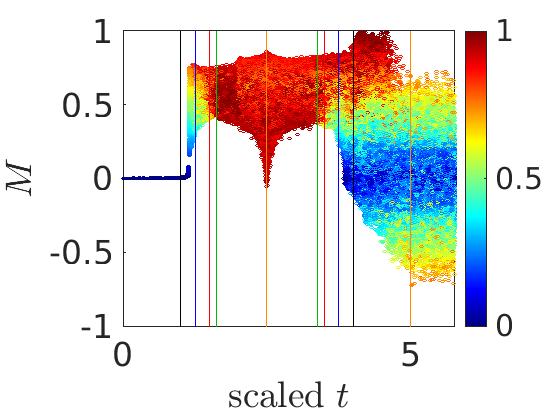} 
\\
\includegraphics[width=4cm]{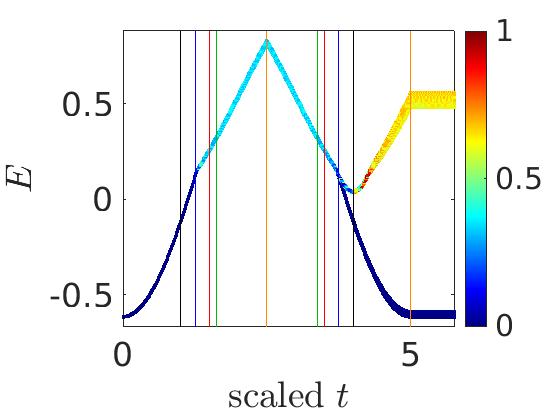} 
\includegraphics[width=4cm]{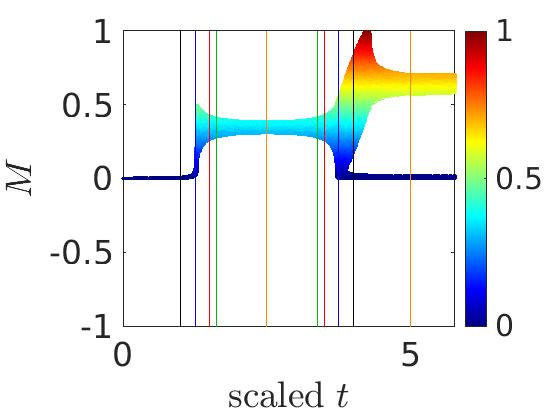} 

\caption{ 
{\bf Simulations with a  \rmrk{tilted} ring.} 
These are additional panels for \Fig{FigTrimerEnergyVsTimeSC} of the main text.
The parameters are the same as for the left panels there (${\dot{\Phi}{=}5\pi \cdot 10^{-4}}$), 
with added  \rmrk{tilt}  $\epsilon{=}0.1$.
The upper panels are generated with the full Hamiltonian,
while the lower panels use TOA. 
}
\label{FigTrimerEnergyVsTimeSC-App} 
\end{figure}

Still one may insist to adopt TOA for a finite $L$ ring. 
How the results would be in comparison with the correct picture?  
Looking in Fig.3 of \cite{SwallowBrand} we see that the interest there 
is in simple adiabatic shuttling along the upper level, 
during which no bifurcation occurs. In this energy range there 
is no major difference between the TOA and the Bogolyubov versions, 
as we see from looking on the higher levels in \Fig{FigDimerEnergyVsTimeQM}.
But for the scenario that we consider, starting at $n{=}0$, 
the TOA completely fails. Demonstration of this colossal failure 
is provide in \Fig{FigTrimerEnergyVsTimeSC-App}. 
\\

\section{Simulations with a \rmrk{tilted} ring}
\label{ApTilt}

\Fig{FigTrimerEnergyVsTimeSC-App} compares the dynamics that is generated by $\mathcal{H}$ 
with the dynamics that is generated using TOA.
In the TOA Hamiltonian we keep just two momentum orbitals. 
Without  \rmrk{tilt} the TOA Hamiltonian is identical with the ${U=0}$ Hamiltonian, 
and therefore its failure is trivial (not displayed).
We therefore add a \rmrk{tilt}  $\epsilon{\ne}0$ as in \cite{SwallowBrand}.      
We see that the TOA completely fails to reproduce the dynamics. 
\\

\section{Bifurcations}
\label{ApF}

The $(S_x,S_z)$ contour lines of the Hamiltonian \Eq{eHdimer}
are ellipses that are chopped by the circle  ${S_x^2+S_z^2} = (N/2)^2$. 
If the circle is ignored, the minimum is at 
\beq \label{gmin}
(S_x,S_z) \ = \ \left( \frac{\mathcal{E}}{2U_{\perp}}, \frac{\epsilon}{2U_{\parallel}} \right)  
\eeq
In the relay shuttling scenario, as $\mathcal{E}$ is varied, 
a bifurcation takes place at the East pole once this minimum 
enters into the circle. This happens at \Eq{eEcSht}.  
In the adiabatic ejection scenario the relevant bifurcation happens on the bounding circle: 
before the bifurcation we have on the circle one minimum and one maximum;  
after the bifurcation a secondary minimum and an associated saddle point appears.    
In order to find the bifurcation we define the function
\beq 
h(\theta) = H\left(S_x{:=}\frac{N}{2}\cos(\theta),S_z{:=}\frac{N}{2}\sin(\theta)\right)
\eeq
Then we write the equations ${h'(\theta)=0}$ and ${h''(\theta)=0}$, 
for the first and the and second derivatives, 
as required at the bifurcation point.    
The combined equations ${ \sin\theta \ h''(\theta) - \cos \theta\ h'(\theta) = 0 }$  
and ${ \cos \theta \ h''(\theta) + \sin \theta \ h'(\theta) = 0  }$  
are solved to get \Eq{eEcDia}. 
\\

\section{Quantum stability of the condensate}
\label{ApH}

For a frozen value of $\Phi$ we perform simulations whose purpose is 
to monitor the stability of the quantum condensate. 
The interest is in the regime ${ \Phi_{\text{stb}} < \Phi <  \Phi_{\text{dyn}} }$.
In this regime the classical cloud has a piece that is trapped 
on a dynamically stable island, and therefore cannot decay.  
But quantum mechanically the cloud can tunnel through the KAM barriers, 
and therefore has a finite lifetime~$\tau$. 
The survival probability ${P(t)= |\Braket{\Psi(0)}{\Psi(t)}|^2  }$ 
of the condensate has been found for representative values of $\Phi$. 
From that $\tau$ has been extracted.  
In the range of interest, for our choice of parameters,  ${ \tau \sim 90 }$.
The survival amplitude is related to the LDOS via a Fourier transform, 
and therefore one can say that we employ here an LDOS based determination of $\tau$.





\sect{Acknowledgements} 
This research was supported by the Israel Science Foundation (Grant No.518/22).

\end{document}